\documentclass[12pt]{article}
\usepackage{amsmath,amssymb}
\newcommand{\be}{\begin{equation}}
\newcommand{\ee}{\end{equation}}
\newcommand{\bea}{\begin{eqnarray}}
\newcommand{\eea}{\end{eqnarray}}
\newcommand{\ba}{\begin{array}}
\newcommand{\ea}{\end{array}}
\newcommand{\p}[1]{(\ref{#1})}
\newcommand{\lb}[1]{\label{#1}}

\def\bbox{{\,\lower0.9pt\vbox{\hrule \hbox{\vrule height 0.2 cm
\hskip 0.2 cm \vrule height 0.2 cm}\hrule}\,}}
\newcommand{\dsl}{\pa \kern-0.5em /}

\newcommand{\nn}{\nonumber \\}

\begin{document}


\begin{titlepage}

\vfill

\begin{center}
\baselineskip=16pt {\Large\bf Dirac Operator on Complex Manifolds
 and  \break
 Supersymmetric Quantum Mechanics}
\vskip 0.3cm
{\large {\sl }}
\vskip 10.mm
{\bf ~E.~A. Ivanov$^{*, 1}$,
A.~V. Smilga$^{\dagger, 2}$}
 \\
\vskip 1cm
$^*$
Bogoliubov Laboratory of Theoretical Physics \\
JINR, 141980 Dubna, Russia\\

\vspace{6pt}
$^\dagger$
SUBATECH, Universit\'e de Nantes, \\
4 rue Alfred Kastler, BP 20722, Nantes 44307, France;\\
On leave of absence from ITEP, Moscow, Russia

\end{center}
\vfill

\par
\begin{center}
{\bf ABSTRACT}
\end{center}
\begin{quote}

We explore a  simple ${\cal N}=2$  SQM model describing the motion over
complex manifolds in external gauge fields.
The  nilpotent supercharge $Q$  of the model can be interpreted as a (twisted)
exterior holomorphic derivative, such that
the model realizes the twisted  Dolbeault complex. The sum $Q + \bar Q$  can  be
interpreted as the Dirac
operator:  the standard Dirac operator if the manifold is
K\"ahler  and the Dirac operator involving certain particular
extra torsions for a generic complex manifold.
Focusing on the K\"ahler
case, we give new simple physical proofs of the two
mathematical facts: {\it (i)} the equivalence of the twisted Dirac and
twisted Dolbeault complexes and {\it (ii)} the Atiyah-Singer theorem.

\vfill
 \hrule width 5.cm
\vskip 2.mm
\noindent $^1$ eivanov@theor.jinr.ru\\
\noindent $^2$ smilga@subatech.in2p3.fr\\

\end{quote}
\end{titlepage}
\setcounter{equation}{0}
\section{Introduction}

{\it Complexes} are
the algebraic objects associated with
smooth manifolds and studied in differential geometry
\cite{EJH}. The most known is
the {\it de Rham} complex involving the exterior derivative $d$
and the Hermitian-conjugate operator
$d^\dagger = (-1)^
{Dp+D+1} \star d \star $  acting on the space of $p$-forms ($D$ is the dimension of the manifold and
$\star$ is the Hodge duality operator). The operators
$d$ and $d^\dagger$ are nilpotent, while their anticommutator
$\{d,d^\dagger \}$ coincides with the covariant Laplacian acting on the forms.
The other important complexes are the {\it Dolbeault} complex, which is defined
on complex manifolds and involves  holomorphic exterior derivative
 $\partial$ and its hermitian conjugate $\partial^\dagger$,
 and the {\it Dirac} complex associated with the Dirac operator.
The complexes may be {\it twisted} by adding background Abelian or non-Abelian gauge
fields\footnote{ There are two parallel terminological systems: physical and mathematical.
For example,
what a physicist calls {\it Abelian gauge field} is called {\it connection on a line bundle}
by a mathematician. We will mostly use the physical terminology.}.

An important characteristics of all these complexes are their indices. The index of an elliptic
operator
\footnote{Recall that elliptic operators are invertible differential operators
that generalize the Laplace operator (and of course include it as a particular case).} ${\cal O}$
can be defined
if the whole Hilbert space of objects (states) where it acts
can be divided into two subspaces (call them ${\cal H}_L$ and ${\cal H}_R$ )
and there are symmetry operators (commuting with ${\cal O}$) that transform a state from ${\cal H}_L$
  into a state from ${\cal H}_R$ and  a state from ${\cal H}_R$
  into a state from ${\cal H}_L$.

In such case, one can always define nilpotent projections: an operator that brings a state
from  ${\cal H}_L$ into a state from ${\cal H}_R$ and annihilates any state from ${\cal H}_R$,
and its Hermitian conjugate: the operator bringing a state
from ${\cal H}_R$ into ${\cal H}_L$ and annihilating the states from ${\cal H}_L$.
  The anticommutator of these nilpotent projections is a symmetry operator too. In the simplest case,
it coincides with ${\cal O}$.
Then all eigenstates of ${\cal O}$ with nonzero eigenvalues are double degenerate (take an eigenstate from  ${\cal H}_L$
and act upon it by a symmetry operator). It is not true for zero eigenvalues.
 The index is then defined as the difference between the number of states
in the kernel of ${\cal O}$  belonging to ${\cal H}_L$ and such a number for ${\cal H}_R$.

For example, for the de Rham complex, ${\cal O} = -\triangle_{\rm cov}$, and the Hilbert space of all relevant forms
can be divided into the subsets of even and odd forms. The relevant symmetry operators are $d$ and $d^\dagger$.
For the Dirac complex\footnote{The sign is chosen so that the operator ${\cal O}$ is positive-definite with $/\!\!\!\!{\cal D}
= {\cal D}_M \gamma^M$
and Hermitian $\gamma^M$.},
 ${\cal O} = -/\!\!\!\!{\cal D}^2$, the Hilbert space of all spinors can be subdivided
into the left-handed spinors and the right-handed ones. The  symmetry operators are $ /\!\!\!\!{\cal D}$
and $/\!\!\!\!{\cal D} \gamma^{D+1}$ (the index of ${\cal O}$ coinciding with the index of $/\!\!\!\!{\cal D}$
 can be defined only for even-dimensional manifolds where $\gamma^{D+1}$, a multidimensional generalization of $\gamma^5$, can be defined).

The indices have beautiful integral representations. Consider, e.g., the 2-dim Dirac operator
in an external Abelian field on the plane. Its index coincides with the magnetic flux,
 \be
\label{indD2}
I_{/\!\!\!\!{\cal D}} \ =\ \frac 1 {2\pi} \int B \, d^2x \ .
 \ee
The integral representations for all indices were systematically
derived by Atiyah and Singer
\cite{AS1,AS2,AS3,AS4}. In their derivation, they used the so called {\it heat kernel}
method
based on the semiclassical ( small
$\beta$) expansion of the
matrix element $\langle x | \Gamma e^{-\beta {\cal O}}|x  \rangle $, where
$\Gamma$ is the grading operator distinguishing between ${\cal H}_L$ and ${\cal H}_R$,
such that $ \Gamma \Psi = \Psi$ when
$\Psi \in {\cal H}_L$ and $\Gamma \Psi = -\Psi$ when
$\Psi \in {\cal H}_R$. (For good pedagogical explanations, see e.g.  \cite{Zumino1,Zumino2}.)

An interesting, from  the physical viewpoint, modification of this
method is based on the observation that the  indices of elliptic
(Euclidean) operators are  associated via the level crossing picture
 with the {\it anomalies} of certain Minkowski space currents \cite{crp1,crp2,crp3}.
For example, the index (\ref{indD2}) is associated with the
anomalous divergence of the 2-dim axial current $J_M = \bar \psi
\gamma_M \sigma^3 \psi$,
 \be
\label{divD2}
 \partial_M J_M \ =\ \frac 1 {4\pi} \epsilon_{MN} F_{MN} \ .
 \ee
One of the ways to derive (\ref{divD2}) is to regularize the current by the Schwinger splitting
$J_M \rightarrow J_M(\epsilon) = \bar \psi(x+\epsilon) \gamma_M \sigma^3 \psi (x-\epsilon)$
and calculate then the Euclidean fermion propagator in external Abelian field $ \langle \psi (x-\epsilon)
\bar \psi(x+\epsilon) \rangle_A $
 using
the Fock-Schwinger gauge technique \cite{LectQCD}.

For the index (\ref{indD2}), the heat kernel calculation is rather explicit, but it is
much more intricate in more complicated cases of mathematical and physical interest.

Back in 1981, Witten noticed \cite{WitMorse1,WitMorse2} that this set of mathematical problems
has a beautiful physical interpretation: the operators ${\cal O}$ can be viewed of as
Hamiltonians of certain supersymmetric quantum mechanics (SQM) systems, while the nilpotent projections
discussed above are interpreted as supercharges. The index of ${\cal O}$ coincides then with the Witten
index of the corresponding SQM system,
 \be
\label{indWit}
I \ =\ {\rm Tr} \{ \Gamma e^{-\beta H} \} \ =\ {\rm Tr} \{ (-1)^F e^{-\beta H} \} \ ,
 \ee
  where $\beta$ is a parameter having the meaning of inverse temperature and $F$ is an operator that commutes with $H$ and
has even eigenvalues for the states from ${\cal H}_L$ and odd eigenvalues for the states from ${\cal H}_R$.
Physically, $F$ is interpreted as the fermion number.
Due to degeneracy between the excited states in ${\cal H}_L$  and ${\cal H}_R$, the index does not depend on $\beta$.

Now, the r.h.s. of (\ref{indWit}) has a functional integral representation. For small $\beta$, this functional integral
can be evaluated by semiclassical methods. As a result, the Atiyah-Singer integral theorems are reproduced. This program
was carried out in \cite{Alv,FW,Wind} (see also, e.g. \cite{Most1,Most2,Most3}).

In our paper, we concentrate on {\it complex} manifolds and study
the ${\cal N}=2$ SQM model\footnote{Following the convention adopted now by the most practitioners
of SQM, ${\cal N}$ counts
the number of {\it real} supercharges. Thus, the minimal interesting case where supersymmetry (double degeneracy of all excited levels) is
present in the spectrum of the Hamiltonian corresponds to ${\cal N} = 2$.} which, in our opinion, is most appropriate
for calculating the relevant indices. Its classical ${\cal N}=2$ superfield
Lagrangian is a particular case of the general Lagrangian given in \cite{Hull}.
In the K\"ahler case, this SQM model is reduced  to the model
considered in \cite{Alv,FW,Wind}, whereas in a generic complex case its Lagrangian is different. Also, in the K\"ahler case,
our approach differs from the approach in Ref. \cite{Alv,FW,Wind} by the choice of supercharges.
Instead of  the supercharges $/\!\!\!\!{\cal D}$ and
$/\!\!\!\!{\cal D} \gamma^{D+1} $ that realize the supersymmetry algebra for any even-dimensional manifold, we use
the (Hermitian) supercharges
 \be
\label{DS}
i/\!\!\!\!{\cal D} = i \gamma^M {\cal D_M}, \ \ \ \ \ \ \ \ \ \ S=  I^M_{\ N} \gamma^N {\cal D}_M\, ,
 \ee where $I^M_{\ N}$ is the
complex structure tensor satisfying the properties
  \be
\lb{comstruc}
 I_{MN} = -I_{NM}, \ \ \ \ \ \ \  I^M_{\ N} I^N_{\ P} = -\delta^M_P \, .
 \ee
The existence of the supercharge $S$ (such that $S^2 = H$ and $\{S, /\!\!\!\!{\cal D} \} = 0$ ) is
specific for K\"ahler manifolds \cite{Wipf}.
The supercharges (\ref{DS}) are naturally obtained in our superfield framework
as a real and imaginary part of a certain complex nilpotent supercharge.

After fixing the complex geometry notations in Sect. 2, we present our model in Sect. 3. In Sect. 4, we show how this ${\cal N} = 2$
SQM model can, in the K\"ahler case, be completed to the extended  ${\cal N} = 4$ SQM model.
  In Sect. 5, we give a geometric interpretation (\ref{DS}) for the N\"other supercharges derived in Sect. 3.
We also observe that the nilpotent supercharge $/\!\!\!\!{\cal D} + S$
can be interpreted as the (twisted) operator of
the holomorphic exterior derivative.
This allows one to prove, in a rather manifest way,  the known mathematical fact:
{\it  for K\"ahler manifolds, the twisted Dirac complex and the
  twisted Dolbeault complex are equivalent}. Sect. 6 is devoted to
the functional integral  derivation of the Atiyah-Singer theorem for the Dirac operator. The derivation is similar in spirit to the derivation in
Ref. \cite{Alv,FW,Wind}, but we do it in a much more detailed way focusing on the K\"ahler case.

Note that the first e-print version of our paper appeared in archive more than 1.5 years ago \cite{arch}. Since then, it inspired a
few papers where its results were further applied and extended \cite{S4,flux,HRR,FIS}. We will comment on these subsequent developments
in the proper places below. We will also make more precise
 some statements of \cite{arch} and add,
in the last Section, a short account of the new results.

\section{Complex geometry}
\setcounter{equation}0

Let us start with recalling some mathematical facts on complex geometry adapted for immediate use in
the next section where the relevant SQM system will be introduced.

We assume the manifold to be even-dimensional of dimension $D = 2n$ and described by complex coordinates $z^N = (z^j, \bar z^{\bar k})$.
The metric is assumed to have the Hermitian form
 $ds^2 = 2 h_{j\bar k} dz^j d\bar z^{\bar k}$.
In other words\footnote{Such a manifold is not necessarily {\it complex} in the precise mathematical sense. The
genuine complex manifold is required to be divisible in several charts
such that
the metric has the  form (\ref{metrcomp}) in each chart
{\it and} the coordinates $(z^j, \bar z^{\bar k})$ in such different overlapping charts are expressed through each other
by means of holomorphic functions. Thus, $S^4$ (in contrast to $S^2$) is not a complex manifold even though
its metric can be represented as
in Eq.(\ref{metrcomp}) in both the northern and the southern hemispheres. Sill notice that the requirement for the metric to
be representable locally in the form (\ref{metrcomp})
is nontrivial and singles out some subset of even-dimensional manifolds.},
 \be
\label{metrcomp}
g_{MN} \ =\ \left( \begin{array}{cc} 0 & h_{j\bar k} \\ h_{k \bar j} & 0 \end{array} \right)  .
 \ee
The covariant derivatives are defined as
\be
\lb{covdevdef}
\nabla_M V^N = \partial_M V^N + \hat{\Gamma}^N_{MP}V^P\, , \ \ \ \ \ \ \
\nabla_M V_N = \partial_M V_N - \hat{\Gamma}^P_{MN}V_P\,
 \ee
(note the order of indices in $\hat{\Gamma}^N_{MP}$, $\hat{\Gamma}^P_{MN}$).
 They involves generically the {\it affine connections}
\be
\hat{\Gamma}^M_{NK} = \Gamma^M_{NK} + \frac{1}{2}g^{ML}C_{LNK}\,, \lb{genGamma}
\ee
where $\Gamma^M_{NK}$ are the standard Christoffel symbols for the metric $g_{MN}$ and $C_{LNK}$ is the
totally antisymmetric torsion tensor.
We will also use  the notation
\be
\label{nabla}
\nabla \psi{\,}^M = \dot{\psi}{\,}^M + \dot z^N \hat{\Gamma}^M_{N L}\psi{\,}^L\,.
\ee

In the following, we will stick to a {\it special} form of the torsion tensor with the nonvanishing components
\be
\lb{Cikl}
C_{j k\bar l} = \partial_{k} h_{j\bar l} - \partial_{j} h_{k\bar l} \, , \quad \quad
C_{\bar j \bar k  l} =  (C_{j k \bar l})^* = \partial_{\bar k} h_{l \bar j} - \partial_{\bar j} h_{l \bar k} \ .
\ee
(and those obtained from them by the cyclic  permutation of indices). In real notations, this can be represented as
\cite{Hull}
 \be
\label{Creal}
C_{JKL} = I_J^{\ P} I_K^{\ R} I_L^{\ T}  \left( \nabla_P I_{RT} +
\nabla_R I_{TP} + \nabla_T I_{PR} \right)
  \ee

The connection (\ref{genGamma}) with the torsion (\ref{Cikl}, \ref{Creal}) is known to mathematicians as
{\it Bismut connection} \cite{Bismut} (see also \cite{Mavr} )
defined as a connection with respect to which both the metric tensor
$g_{MN}$ and
the complex structure tensor $I_{MN}$ are covariantly constant while the torsion tensor $C_{JKL}$ is completely
antisymmetric   

 The torsion \p{Creal} has, generically, a non-zero curl,
$\partial_{[M} C_{JKL]} \neq 0$
\footnote{This is  in contrast to Ref. \cite{Mavr,Bismut}
 where the Atiyah-Singer theorem on the manifolds
involving extra curl-free torsion was considered.}.

    Non-vanishing components of $\Gamma_{M, NP}$ and $\hat\Gamma_{M, NP}$ are
\bea
&&\Gamma_{\bar m,  n p} = \frac{1}{2}\left(\partial_n h_{p\bar m} + \partial_{p}h_{n\bar m}\right), \quad
\Gamma_{m, n \bar p} = \Gamma_{m, \bar p n} = \frac{1}{2} C_{mn \bar p} , \quad (\mbox{and c.c.})\,, \nn
&& \hat{\Gamma}_{\bar m, n p } = \partial_p h_{n\bar m}\,, \quad \hat\Gamma_{m, n \bar p} =
C_{mn \bar p}\,,
\quad \hat\Gamma_{m, \bar p n} = 0\,, \quad  (\mbox{and c.c.})\,. \lb{nonz}
\eea

The last identity means that, after a parallel transport with the Bismut affine connections,
an (anti)holomorphic vector is transformed into an (anti)holomorphic vector: the holonomy group is $U(n)$ rather than $SO(2n)$.
As is seen from the first line in (\ref{nonz}), this property does not hold for usual torsionless covariant derivatives
for a generic complex manifold.

Let us introduce complex vielbeins as
\bea
&& e_k^a e^{\bar a}_{\bar i} = h_{k\bar i}\,, \quad e^k_a e^{\bar i}_{\bar a} = h^{\bar i k}\,, \;\;
h^{\bar i k}h_{k \bar j} = \delta^{\bar i}_{\bar j}\,, \;\; h_{k \bar j}h^{\bar j l} = \delta^l_k\,, \;\;  \lb{viel} \\
&& e_k^a\,e_a^j = \delta^j_k\,,\;\;e^k_a\,e^b_k = \delta^b_a\,,\;\;e_{\bar k}^{\bar a}\,e_{\bar a}^{\bar j} = \delta^{\bar j}_{\bar k}\,,\;\;
e^{\bar k}_{\bar a}\,e^{\bar b}_{\bar k} = \delta^{\bar b}_{\bar a}\,. \lb{ort}
\eea

The nonzero components of the standard spin connections
$$\Omega_{M, AB} \ =\ e_{AN} (\partial_M e^N_B + \Gamma^N_{MK} e^K_B) $$
are
 \bea
\lb{conn}
\Omega_{j, \bar b a} = - \Omega_{j, a \bar b} = e_{p}^b (\partial_j
 e^p_a + \Gamma^p_{jk} e^k_a ), \ \ \ \ \ \ \
\Omega_{j, \bar a \bar b} \ =\ \frac 12 e^{\bar s}_{\bar a} e^{\bar k}_{\bar b} C_{j \bar s \bar k}
 \eea
and complex conjugated $\bar \Omega_{\bar j, b \bar a}, \ \ \bar \Omega_{\bar j, ab}$, with $C_{j \bar k \bar s}$ defined in (\ref{Cikl}).

When the torsion is present, one can define a generalized spin connection related to the generalized affine
connection $\hat{\Gamma}_{ML}^N$:
\be
\hat \Omega_{M, AB} = e_{AN} (\partial_M e^N_B + \hat\Gamma^N_{MK} e^K_B) = \Omega_{M, AB} - \frac{1}{2}e_A^L e_B^K C_{MLK}\,.\lb{hatOm}
\ee
The nonzero components of $\hat\Omega$ are
\bea
&& \hat{\Omega}_{i, \bar b a} = -\hat{\Omega}_{i, a\bar b}= e^b_k \partial_ie^k_a + e^{\bar t}_{\bar b}
 e^k_a \hat{\Gamma}_{\bar t, i k} = e^{k}_{a}\partial_k e^{b}_{i} - e^{k}_{a}\partial_i e^{b}_{k}
+ e^k_a e^{\bar t}_{\bar b} e^c_i \partial_k e^{\bar c}_{\bar t}\,, \nn
&& \hat{\bar\Omega}_{\bar i, b \bar a} = (\hat{\Omega}_{i, \bar b a})^*\,,
\eea
while the components $\hat \Omega_{j, \bar a \bar b}$ and  $\hat \Omega_{\bar j,  a  b}$ vanish.

The vielbeins and the generalized spin connection satisfy the Maurer-Cartan structure equation
 \be
\lb{Cartan}
de_A + \hat \Omega_{AB} \wedge e_B = T_A \, ,
 \ee
where
$$ e_A = e_{AM}\, dx^M\, , \ \ \  \hat \Omega_{AB} = \hat \Omega_{M, AB} \, dx^M \, , \ \ \
T_A  = \frac 12 e^M_A C_{MNP}\, dx^N \wedge dx^P \, .$$
The  Maurer-Cartan equation for the standard torsion-free spin connection is
\be
\lb{Cartan1}
de_A + \Omega_{AB} \wedge e_B = 0\,.
\ee
The equations \p{Cartan} and \p{Cartan1} are equivalent, as can be checked using the relation \p{hatOm}.
For the Hermitian metric (\ref{metrcomp}), with the torsion defined in (\ref{Cikl}),
these equations imply the identity
   \be
\partial_{[k}e^a_{l]} - e^{\bar i}_{\bar a}\, e^d_{[k}\,\partial{}_{l]}e^{\bar d}_{\bar i}=
\frac{1}{2}\,e^{\bar j}_{\bar a}\,C_{lk\bar j} \quad (\mbox{and c.c.})\,.\lb{kconstr-e2}
   \ee

For K\"ahler manifolds, the metric (\ref{metrcomp}) is derived from the K\"ahler potential,
\be
h_{j\bar k}(z, \bar z) = \partial_j\partial_{\bar k} K(z, \bar z)\,.
\ee
In this case
\be
\partial_{\bar l} h_{j\bar k} - \partial_{\bar k} h_{j\bar l} = \partial_{k} h_{j\bar l} -
\partial_{j} h_{k\bar l} = 0\,,\lb{cond-g}
\ee
and, as a result,
\be
C_{MNK} = 0 \; \Rightarrow \; \hat\Gamma^M_{NK}  = \Gamma^M_{NK}\,.
\ee
 The only nonvanishing components of $\Gamma_{M,NP}$
in the K\"ahler case are:
\be
\Gamma_{\bar m, np}= \partial_n h_{p\bar m}\,, \quad \Gamma_{m, \bar n \bar p} = (\Gamma_{\bar m, np})^*
= \partial_{\bar n}h_{m\bar p}\,.
\ee
 The expressions for the non-vanishing components of the spin connections are also greatly simplified,
 \be
\hat {\Omega}_{j, \bar b a} =  \Omega_{j, \bar b a} =
e^{\bar k}_{\bar b}\partial_j e^{\bar a}_{\bar k} \stackrel {\rm def}=   \omega_{j, \bar b a}\,,
\quad \hat {\bar\Omega}_{\bar j, b \bar a} =  \bar\Omega_{\bar j, b \bar a }  =
e^k_b\partial_{\bar j}e^a_k  \stackrel {\rm def}=  \bar\omega_{\bar j, b \bar a }\, . \lb{omkal}
\ee

In the K\"ahler case, the structure equation
(\ref{kconstr-e2}) acquires the form
 \be
\lb{kconstr-e}
\partial_{[k}e^a_{l]} - e^{\bar i}_{\bar a}\, e^d_{[k}\,\partial{}_{l]}e^{\bar d}_{\bar i}= 0 \quad (\mbox{and c.c.})\,,
\ee
which is now a non-trivial constraint on the vielbeins $e^a_l, e^{\bar a}_{\bar l}$ ( equivalent to the constraint \p{cond-g}
on the metric).
The only non-vanishing components of the K\"ahler Riemann tensor are
\bea
\lb{Riemann}
R_{j\bar k, l \bar t} =  \partial_j\partial_{\bar k} h_{l \bar t} - h_{p\bar s}\,\Gamma^p_{j\,l}\Gamma^{\bar s}_{\bar k\,\bar t} =
\partial_j\partial_{\bar k} h_{l \bar t} - h^{\bar s n}\,(\partial_j h_{l \bar s})\,(\partial_{\bar k}h_{n \bar t}) \nn
= e_l^{a} e_{\bar t}^{\bar b}
\left( \partial_j \bar \omega_{\bar k, a \bar b} + \partial_{\bar k} \omega_{j, \bar b a} +
\omega_{j, \bar b d} \bar \omega_{\bar k, a \bar d} - \omega_{j, \bar d a} \bar \omega_{\bar k, d \bar b} \right) \, .
\eea

Finally, note the useful generic  relations:
\be
 \Omega_{i, \bar a b} = \omega_{i, \bar a b} + \frac 12 e^l_b e^{\bar t}_{\bar a}C_{il\bar t}\,, \quad
\hat \Omega_{i, \bar a b} = \omega_{i, \bar a b} +  e^l_b e^{\bar t}_{\bar a}C_{il\bar t}\,, \lb{Relimp}
\ee
where the expressions $\omega_{i, \bar a b}$ coincide by form with those defined in (\ref{omkal}). Note that
the objects $\omega_{i, \bar a b}$  can be given a geometric interpretation even in a non-K\"ahler case. They coincide with
 the appropriate components of
a generalized spin connection associated with $\tilde{\Gamma}^{M}_{NK} = \Gamma^{M}_{NK} - \frac{1}{2} g^{MT} C_{TNK}\,$.

\section{${\cal N}=2$ SQM model }
\setcounter{equation}0

We formulate a general complex ${\cal N}=2, d=1$ SQM sigma model
in terms of $2n$ mutually conjugated chiral and anti-chiral superfields
$Z^j(t_L,\theta), \bar Z^{\bar{j}}(t_R, \bar\theta)\, (j,\bar j = 1, \cdots n)\,$,
\be
\bar D Z{\,}^j(t_L,\theta) = D \bar Z{\,}^{\bar{j}}(t_R, \bar\theta) = 0\,,
\ee
where
\be
D = \frac{\partial}{\partial \theta} - i\bar\theta \partial_t\,,
\bar D = -\frac{\partial}{\partial \bar\theta} + i\theta \partial_t\,,\quad \{D, \bar D \} =2i \partial_t, \; \;
t_L = t - i\theta\bar\theta, t_R = t + i\theta\bar\theta\,.
\ee
The basic superfields have the following component expansion
\be
Z{\,}^j = z{\,}^j +\sqrt{2}\,\theta \psi{\,}^j - i\theta\bar\theta \,\dot{z}{\,}^j\,, \quad \bar Z{\,}^{\bar{j}} = \bar z{\,}^{\bar{i}}
-\sqrt{2}\,\bar\theta \bar\psi{\,}^{\bar{j}} +
i\theta\bar\theta \dot{\bar{z}}{\,}^{\bar{j}}\,.
\ee
The ${\cal N}=2$ transformation properties of the component fields are as follows:
\bea
&& \delta z{\,}^j = -\sqrt{2} \epsilon \psi^j\,, \quad \delta \psi{\,}^j = \sqrt{2}i\,\bar\epsilon\,\dot{z}{\,}^j\,, \nn
&& \delta \bar z{\,}^{\bar j} = \sqrt{2}\bar\epsilon\, \bar\psi{\,}^{\bar j}\,, \quad
\delta\bar\psi{\,}^{\bar j} = -\sqrt{2} i\,\epsilon\,\dot{\bar{z}}{\,}^{\bar j}\,.\lb{transfcomp}
\eea

The superfield action we start with reads
\bea
\lb{start}
&& S = \int dt d^2\theta \left({\cal L}_\sigma  + {\cal L}_{gauge}\right), \nn
&&
{\cal L}_\sigma = -\frac{1}{4} h_{i\bar j}(Z, \bar Z)\, D Z{\,}^i \bar D\bar Z{\,}^{\bar j}\,, \quad {\cal L}_{gauge} =
\, W(Z, \bar Z)\,.
\eea

Here, $ h_{i\bar j}(Z, \bar Z)$ and $W(Z, \bar Z)$ are unconstrained functions of the  superfields. In general \cite{Hull}, one can
add to ${\cal L}$ the terms
\be
\sim {\cal B}_{ik}(Z, \bar Z)D Z{\,}^{i} D Z{\,}^{k} + c.c.\,. \lb{missed}
\ee
and also the terms involving higher antisymmetric even-dimensional tensors ${\cal B}_{ijkl}$, etc.
These additional terms do not change the target space metric in the component action and affect only fermionic terms (introducing
some extra non-zero components of the torsion). In this paper, we concentrate on the model where all these tensors vanish.
 The models with nonzero ${\cal B}$ were discussed in a recent paper \cite{FIS}.

The K\"ahler case corresponds to the choice
\be
\lb{Kahlmetr}
 h_{j\bar k}(Z, \bar Z) = \partial_{j}\partial_{\bar k}K(Z, \bar Z)\,,
\ee
where the K\"ahler potential $K(Z, \bar Z)$ is an arbitrary real function of the superfields
\footnote{For $\mathbb{CP}^1$ and with the restriction $K = W$, this SQM model
  was earlier considered
at the classical (component and superfield)  and quantum levels
 in Refs.\cite{quantCP11,quantCP12,quantCP13}.}.

The component form of the full action is
\bea
&&S \equiv  \int dt \left(L_\sigma + L_{gauge}\right)= \int dt \Big\{ h_{j\bar k} \left[\dot{z}{\,}^j\dot{\bar z}{\,}^{\bar k}
+\frac{i}{2} \left( \psi{\,}^j \dot{\bar{\psi}}{\,}^{\bar k} - \dot\psi{\,}^j \bar{\psi}{\,}^{\bar k}\right)\right] \nn
&& \qquad -\, \frac{i}{2}\left[\left(2\partial_j h_{t\bar k} - \partial_t h_{j\bar k}\right)\dot z{\,}^t
-  \left(2\partial_{\bar k} h_{j \bar t} -
\partial_{\bar t} h_{j\bar k}\right)\dot{\bar z}{\,}^{\bar t}\right]\psi{\,}^j \bar\psi{\, }^{\bar k}
+ \left(\partial_t\partial_{\bar l} h_{j\bar k}\right) \psi{\,}^t\psi{\,}^j \bar\psi{\,}^{\bar l}\bar\psi{\,}^{\bar k} \nn
&& \qquad +\, 2 \partial_j\partial_{\bar k}W\,\psi{\,}^j\bar\psi{\,}^{\bar k} -  i \left(\partial_j W\dot z{\,}^j
- \partial_{\bar j}W \dot{\bar z}{\,}^{\bar j}\right) \Big\}\,. \lb{genAct}
\eea

The geometric meaning of the different
terms in the  ``sigma-model''  part  $L_\sigma$ of the  Lagrangian in
(\ref{genAct}) can be clarified, if rewriting it in the following form
\bea
L_\sigma = \frac{1}{2}\left[ g_{MN}\,\dot z{\,}^M \dot z{\,}^N + ig_{MN}\,\psi^M\nabla \psi^N
- \frac{1}{3!}\,\partial_P C_{MNT}\,\psi^P\psi^M\psi^N\psi^T \right], \lb{1comp}
\eea
where the metric $g_{MN}$ is written in (\ref{metrcomp}), the covariant derivative
$\nabla \psi^N$ was defined in (\ref{nabla}) and the torsion tensor $C_{MNT}$ in
(\ref{Cikl}), (\ref{Creal}).

The Lagrangian $L_{gauge}$ (the last line in \p{genAct}) describes the interactions with the Abelian gauge field
$A_M = (-i\partial_j W, i\partial_{\bar j} W) \equiv  I_M^{\ N} \partial_N W$,
the double derivative $F_{j \bar k} = -F_{\bar k j} =  2i\partial_j \partial_{\bar k}W$ being the magnetic field strength.
This Lagrangian can also be rewritten in a form
analogous to \p{1comp}
\be
L_{gauge} =
A_M \dot{Z}^M  - \frac{i}{2}\, F_{MN}\psi^M\psi^N\,. \lb{gaugComp}
\ee
The prepotential $W(Z, \bar Z)$ is an arbitrary function.
  A particularly clever choice  is $W \propto \ln \det h$ (see
Eq.(\ref{choiceW}) below). The corresponding bundle $(-i\partial_j W, i\partial_{\bar j} W)$
is called {\it canonical} by mathematicians.

The fermion variables $\psi^j, \bar\psi^{\bar k}$ are not canonically conjugated, their Poisson bracket being
equal to $-ih^{\bar k j}$.
It is convenient to introduce the canonically conjugated fermionic fields with the tangent space indices,
\be
\psi^a = e^a_j\,\psi^j\,, \quad \bar\psi{\,}^{\bar b} = e^{\bar b}_{\bar k}\bar\psi{\,}^{\bar k}\,,
\ee
such that all the variables have the
canonical Poisson brackets,
\be
\label{basicPB}
\{z{\,}^j, P_k \}_{PB} = \delta^j_k\,, \quad \{\bar z{\,}^{\bar j}, P_{\bar k} \}_{PB} = \delta^{\bar j}_{\bar k}\,, \quad
\{\psi{\,}^a, \bar\psi{\,}^{\bar b} \}_{PB} = - i \delta^{a\bar b}\,.
\ee

Then, using the invariance of the Lagrangian in \p{genAct}  under the transformations
\p{transfcomp} (modulo a total time derivative),
it is easy to compute the corresponding canonical supercharges
 \bea
&& Q = \sqrt{2}\left[ \Pi_k e^k_a\psi{\,}^a -
i \psi{\,}^b\psi{\,}^d\bar\psi{\,}^{\bar a}(e^k_b\partial_{[k} e^a_{l]} e^l_d)\right], \nn
&& \bar Q= \sqrt{2}\left[ \bar\Pi_{\bar k} e^{\bar k}_{\bar a}\bar\psi{\,}^{\bar a} -
i \bar\psi{\,}^{\bar c}\bar\psi{\,}^{\bar a}\psi{\,}^d(e^{\bar k}_{\bar c}\partial_{[\bar k}e^{\bar d}_{\bar l]}
e^{\bar l}_{\bar a})\right] \, ,\lb{QQmod0}
  \eea
where
\be
\Pi_k = P_{k}   + i \, \partial_k W\,, \quad \bar\Pi_{\bar k} =
P_{\bar k} - i \,\partial_{\bar k} W\,,
  \ee
and $P_k, P_{\bar k}$ are the canonical momenta.

Using the definitions (\ref{conn}), (\ref{Cikl}) and the relations \p{kconstr-e2}, \p{Relimp}, these supercharges can be brought
in a more suggestive geometric form
\bea
Q = \sqrt{2}\left[ \Pi_k - i \bar\psi{\,}^{\bar a} \psi{\,}^b\,\Omega_{k, \bar a b}
\right]e^k_c\psi{\,}^c , \quad
\bar Q= \sqrt{2}e^{\bar k}_{\bar c}\bar\psi{\,}^{\bar c}\left[ \bar\Pi_{\bar k}  +
i \bar\psi{\,}^{\bar a} \psi{\,}^{d}\,\bar{\Omega}_{\bar k, d \bar a}\right].
\lb{QQmod2}
\eea

It should be pointed out that the 3-fermionic terms in these supercharges contain the spin connections $\Omega_{k, \bar a b},
\bar{\Omega}_{\bar k, d \bar a}$ corresponding to the standard symmetric Christoffel symbols $\Gamma^N_{MK}$ and defined by the relations
\p{conn}, \p{Relimp}.
\footnote{On the other hand, the supercharges involve only the holomorphic components of the spin connections
and do not depend on non-holomorphic components
$\Omega_{k, \bar a \bar b}$, $\Omega_{\bar k,  a b}$. See also Eq.(\ref{Qreal}) below.}

Using (\ref{basicPB}),
it is easy to find
\bea
\{Q, \bar Q\}_{PB} = -2i H_{cl}\,,\;\{Q, Q\}_{PB} = \{\bar Q, \bar Q\}_{PB} = [Q, H_{cl}]_{PB} = [\bar Q, H_{cl}]_{PB} = 0.\lb{PBsusy}
\eea
 The canonical classical Hamiltonian  $H_{cl}$ can be represented in the following compact form:
\bea
&&H_{cl} = h^{\bar k j}\left( \Pi_{j} - i \hat{\Omega}_{j, \bar b a}\,\bar\psi{\,}^{\bar b}\psi{\,}^a\right)
\left(\bar\Pi_{\bar k} + i \hat{\bar\Omega}_{\bar k, c\bar d}\,\bar\psi{\,}^{\bar d}\psi{\,}^c\right)
 \nn
&&  \quad -\, 2 e^j_a e^{\bar k}_{\bar b}\partial_j\partial_{\bar k} W \psi{\,}^a\bar\psi{\,}^{\bar b} -
e^t_a e^j_c e^{\bar l}_{\bar b} e^{\bar k}_{\bar d}(\partial_t\partial_{\bar l}{\,}h_{j\bar k})\psi{\,}^a\psi{\,}^c
\bar\psi{\,}^{\bar b}\bar\psi{\,}^{\bar d}\,. \lb{Hclass}
\eea

 It is interesting to note that, in the generic case, the
spin connections entering the supercharges \p{QQmod2} and the classical
hamiltonian \p{Hclass} do not coincide with each other: they differ
by a term proportional to the torsion.

In the  K\"ahler case,  when the torsion is vanishing, the situation simplifies.
Both the supercharges and the Hamiltonian are
expressed through the same connections \p{omkal}.
In addition, the last four-fermionic term in \p{Hclass}  vanishes.

We thus obtain
\bea
Q^{\rm K\ddot{a}hl} = \sqrt{2}\left[ \Pi_k - i \bar\psi{\,}^{\bar a} \psi{\,}^b\,\omega_{k, \bar a b}
\right]e^k_c\psi{\,}^c , \quad
\bar Q^{\rm K\ddot{a}hl} = \sqrt{2}e^{\bar k}_{\bar c}\bar\psi{\,}^{\bar c}\left[ \bar\Pi_{\bar k}  +
i \bar\psi{\,}^{\bar a} \psi{\,}^{d}\,\bar{\omega}_{\bar k, d \bar a}\right]
\lb{QclKahl}
\eea
and
\bea
H_{cl}^{\rm K\ddot{a}hl} =
h^{\bar k j}\left( \Pi_{j} - i {\omega}_{j, \bar b a}\,\bar\psi{\,}^{\bar b}\psi{\,}^a\right)
\left(\bar\Pi_{\bar k} + i {\bar\omega}_{\bar k, c\bar d}\,\bar\psi{\,}^{\bar d}\psi{\,}^c\right)
 -2  e^j_a e^{\bar k}_{\bar b}\, \partial_j\partial_{\bar k} W \psi{\,}^a\bar\psi{\,}^{\bar b} \,. \lb{HclKahl}
\eea

The expression for the Lagrangian also simplifies a lot in the K\"ahler case.
The four-fermionic term in \p{genAct}, \p{1comp} vanishes.
The remaining terms in
$L_\sigma$ can be presented as
 \be
\lb{LKahler}
L_\sigma^{\rm K\ddot{a}hl} \ =\ h_{j \bar k} \dot z^j \dot {\bar z}^k  +
\frac i2  (\psi^a \dot{\bar\psi}^{\bar a} - \dot{\psi}^a \bar\psi^a )
+ i \left(\dot{\bar z}^{\bar k} \bar\omega_{\bar k, a \bar b}- \dot{z}^k \omega_{k,\bar b a} \right)  \psi^a \bar \psi^{\bar b} \, .
 \ee

Let us now turn to quantum theory. The Poisson brackets (\ref{basicPB}) are replaced
by the (anti)commutators:
\be
[z^j, P_k] = i\delta^j_k\,, \; [\bar z^{\bar j}, P_{\bar k}] = i\delta^{\bar j}_{\bar k}\,,
\; \{\psi^a, \bar\psi^{\bar b}\} = \delta^{a\bar b}\,.
\ee
As is well known, there exist, generically, many different quantum theories corresponding to a given
classical one,  due to ordering ambiguities.  To make a selection,  we require that the supersymmetry algebra
(\ref{PBsusy}) remains intact at the quantum level and that $Q_{qu}$ and $\bar Q_{qu}$ are Hermitian conjugate to each other.
 As was noticed in \cite{howto}, these two requirements can be simultaneously fulfilled only provided that the classical expressions
 for the {\it supercharges} are Weyl-ordered in the quantum case.
 The correct expression for the quantum Hamiltonian is obtained as the anticommutator of the Weyl-ordered $Q_{qu}$ and $\bar Q_{qu}$.
Note that this correct quantum Hamiltonian {\it does}
not coincide with the operator obtained through  Weyl-ordering of the classical Hamiltonian
defined by the relations \p{PBsusy}.

We thus obtain
\bea
&& Q^{\rm flat} = \frac{1}{\sqrt{2}}\left[\{\Pi_k,  e^k_a\} \psi{\,}^a
 + i \{\psi{\,}^b\psi{\,}^d, \bar\psi{\,}^{\bar a}\}e^k_b\,\Omega_{k, \bar a d}\right], \nn
&& \bar Q^{\rm flat} = \frac{1}{\sqrt{2}}\left[ \{\bar\Pi_{\bar k}, e^{\bar k}_{\bar a}\}\bar\psi{\,}^{\bar a} +
i \{\bar\psi{\,}^{\bar c}\bar\psi{\,}^{\bar a}, \psi{\,}^d\} e^{\bar k}_{\bar c}\,\bar\Omega_{\bar k,d \bar a}\right].\lb{QQmod3}
\eea

These Weyl-ordered supercharges were dubbed ``flat'' because
they act on the wave functions normalized by the condition \cite{howto}
\be
\label{normflat}
\int \prod_k dz^k d\bar z^{\bar k} \prod_a d\psi^a d\bar\psi^{\bar a} \exp\{\bar \psi^{\bar a} \psi^a\} \, \bar \Psi(z^k, \bar z^{\bar k},
\bar \psi^{\bar a})
 \Psi(\bar z^{\bar k}, z^k, \psi^a) \ =\ 1\,
 \ee
with the flat Hilbert space measure.
 In particular, it is straightforward to see that the Weyl-ordered supercharges $Q$ and $\bar Q$
are Hermitian-conjugate to each other with respect
to such flat inner product\footnote{The same concerns the fermion operators  $\psi^a$ and $\bar\psi{\,}^{\bar a} = \frac{\partial}{\partial \psi^a}$.
They are Hermitian-conjugated with a particular Berezin integration measure in \p{normflat}, \p{normcov} involving the factor
$ \exp\{\bar \psi^{\bar a} \psi^a\}$.}.

It is more natural, however, to deal with the covariant supercharges $Q^{\rm cov}, \bar Q^{\rm cov}$ which act on the Hilbert space
in which the inner product is defined with the covariant integration measure
 \be
\label{normcov}
\int \, \prod_k  \ dz^k d\bar z^{\bar k}\,\det h  \prod_a d\psi^a d\bar\psi^{\bar a} \exp\{\bar \psi^{\bar a} \psi^a\} \,
\bar \Psi(z^k, \bar z^{\bar k}, \bar \psi^{\bar a})
 \Psi(\bar z^{\bar k}, z^k, \psi^a) \ =\ 1
 \ee
(note that $\det h \propto \sqrt{\det g}$). They are related to the flat supercharges by a similarity
 transformation
\be
\lb{similar}
(\,Q^{\rm cov}, \bar Q^{\rm cov}\,) = (\det h)^{-1/2}\,(\,Q^{\rm flat}, \bar Q^{\rm flat}\,)(\det h)^{1/2}\,,
\ee
which yields the expressions:
\bea
&& Q^{\rm cov} = \sqrt{2}\psi^c e^k_c\left[\Pi_k -\frac{i}{2} \partial_k  (\ln \det \bar e)
+ i \psi{\,}^b \bar\psi{\,}^{\bar a}\,\Omega_{k, \bar a b}\right] \nn
&& \bar Q^{\rm cov} = \sqrt{2}\bar\psi{\,}^{\bar c} e^{\bar k}_{\bar c}
\left[\bar\Pi_{\bar k} -\frac{i}{2} \partial_{\bar k} (\ln \det e)
+ i \bar\psi{\,}^{\bar b} \psi{\,}^{a}\,\bar{\Omega}_{\bar k, a \bar b}\right]. \lb{Qcovgen}
\eea
Here,
\be
\lb{PidW}
\Pi_k = -i\left(\frac{\partial}{\partial z^k} - \,\partial_k W\right), \quad
\bar\Pi_{\bar k} = -i \left(\frac{\partial}{\partial \bar z{\,}^{\bar k}} + \,\partial_{\bar k} W\right).
\ee
The supercharges obey the relations of the ${\cal N}=2, d=1$ Poincar\'e superalgebra
\bea
\{Q^{\rm cov}, \bar Q^{\rm cov}\} = 2 H_{qu}^{\rm cov}\,,\;\{Q^{\rm cov}, Q^{\rm cov}\} = \{\bar Q^{\rm cov}, \bar Q^{\rm cov}\}
= [Q^{\rm cov}, H_{qu}^{\rm cov}] =
[\bar Q^{\rm cov}, H^{\rm cov}_{qu}] = 0\,.\nonumber
\eea

The expression for  the quantum Hamiltonian $H^{\rm cov}_{qu}\,$ can be obtained
in two ways: ${\rm (i)}$ By directly calculating the anticommutator of quantum supercharges \p{Qcovgen}
or ${\rm (ii)}$ by
Weyl-ordering the Gr\"onewold-Moyal bracket \cite{Groen,Moyal} of the classical supercharges \p{QQmod2} and performing
then the similarity transformation, like in \p{similar}.

 We obtain
 \bea
H^{\rm cov}_{qu} &=& - \frac 12 \triangle^{\rm cov} + \ \frac 18 \left (R - \frac 12 h^{\bar k j}h^{\bar l t}h^{\bar i n}
 C_{j\,t\, \bar i}\,C_{\bar k\,\bar l\, n}\right) \nn
&& -\, 2 \langle \psi^a \bar \psi^{\bar b} \rangle\,e^k_a e^{\bar l}_{\bar b}\partial_k\partial_{\bar l} W
- \langle \psi^a \psi^c \bar \psi^{\bar b} \bar \psi^{\bar d} \rangle \,
 e^t_a e^j_c e^{\bar l}_{\bar b} e^{\bar k}_{\bar d} \, (\partial_t\partial_{\bar l}{\,}h_{j\bar k})\,.  \lb{kvantH}
\eea
Here, $\langle \ldots \rangle$ denotes the Weyl-ordered products of fermions, $R$ is the standard scalar curvature
of the metric $h_{j\bar k}$, and $\triangle^{\rm cov}$ is the covariant Laplacian calculated with the
``hatted'' affine connections in \p{nonz} and including also the
(hatted) spin connections,
 \be
-\triangle^{\rm cov} \ =\ h^{\bar k j} \left( {\cal P}_j {\bar {\cal P}}_{\bar k} +
i \hat \Gamma^{\bar q}_{j \bar k} {\bar {\cal P}}_{\bar q}
+  {\bar {\cal P}}_{\bar k} {\cal P}_j + i \hat \Gamma^{s}_{\bar k j} {{\cal P}}_s \right),
 \ee
where ${\cal P}_j = \Pi_j + i \hat \Omega_{j, \bar b a} \langle\psi^a \bar \psi^{\bar b}  \rangle $ and
  ${\bar {\cal P}}_{\bar k} = \bar \Pi_{\bar k} - i \hat {\bar \Omega}_{\bar k,  a \bar b}
\langle \psi^a \bar \psi^{\bar b} \rangle \,$. Note
  that the scalar curvature $R$ is related to its ``hatted'' counterpart $\hat R$
associated with the non-symmetric affine connection $\hat \Gamma$ by
the simple formula
$$\hat R = R - \frac 14 C_{MNP}C^{MNP} = R - \frac 32 h^{\bar k j}h^{\bar l t}h^{\bar i n}
 C_{j\,t\, \bar i}\,C_{\bar k\,\bar l\, n}\,.$$

 In the K\"ahler case, the expression for the quantum Hamiltonian greatly simplifies:
 \bea
\lb{kvantHKahl}
H^{\rm cov}_{\rm K\ddot{a}hl} \ =\   - \frac 12 \triangle^{\rm cov} + \frac R8
-\, 2 \langle \psi^a \bar \psi^{\bar b} \rangle\,e^k_a e^{\bar l}_{\bar b}\partial_k\partial_{\bar l} W\,,
\eea
where now $-\triangle^{\rm cov} = h^{\bar k j} \left( {\cal P}_j {\bar {\cal P}}_{\bar k} +  {\bar {\cal P}}_{\bar k} {\cal P}_j \right)$
and $\hat \Omega, \,\hat{\bar \Omega} $ are reduced to $\Omega, \,\bar \Omega = \omega, \,\bar\omega$ according to the relations \p{omkal}.

An important remark is to the point here. The Lagrangian \p{LKahler} can also be expressed through real variables,
 \bea
\label{N=1}
L = \frac{ g_{MN}}{2}\left[\dot z{\,}^M \dot z{\,}^N + i\psi^M\nabla \psi^N \right ]
=   \frac{1}{2}\left[g_{MN}\dot z{\,}^M \dot z{\,}^N  + i\psi^A (\dot \psi^A + \Omega_{M, AB} \dot z^M\psi^B) \right].
 \eea
This Lagrangian is well known \cite{Alv,FW}. It can be (and was) also  considered for a generic (not necessarily complex)
manifold. In a generic case, it is manifestly invariant only under ${\cal N} =1$ supersymmetry transformations
(with a real Grassmann parameter). The corresponding  N\"other supercharge is
 \be
\label{Qreal1}
{\cal Q} \ \sim \ \psi^A e^M_A \left[ P_M -  \frac i2  \Omega_{M, BC} \psi^B \psi^C \right ].
 \ee
 The covariant quantum supercharge (obtained by Weyl-ordering of the classical supercharge and taking
 a correct account of the measure factor as in \p{similar}) is given by the {\it same}
expression, where now $\{\psi^A, \psi^B\} = \delta^{AB} $. It
  can be interpreted as  the Dirac operator. By construction, it is Hermitian with the Hilbert space metric including the factor
$\sqrt{\det g}\,$.

The corresponding quantum Hamiltonian \cite{Zumino1,Zumino2} coincides, up to a proper similarity transformation,
with \p{kvantHKahl} rewritten in real notations.

As was mentioned in the Introduction (and we will return to the discussion of this issue in Sect. 6),
for an even-dimensional
manifold, the second real supercharge
 \be
 \label{Qgamma5}
\tilde {\cal Q} \ =\ 2^{D/2} {\cal Q} \prod_{A=1}^D \, \psi^A
 \ee
associated with $/\!\!\!\!{\cal D} \gamma^{D+1}$ can also be
defined. However, for $D\geq 4$, this second supercharge has nothing to do with the supercharges \p{Qcovgen}.

To recapitulate:
  \begin{itemize}
  \item  For any even-dimensional manifold, the system  (\ref{N=1}) admits two real quantum supercharges
\p{Qreal1} and \p{Qgamma5}.
   \item For any manifold with Hermitian metric \p{metrcomp}, a generically
{\it different} system \p{genAct} involves a {\it different}
pair of  supercharges  \p{Qcovgen}. We will show in Sect. 5 that these supercharges
can be interpreted as an exterior holomorphic derivative and its complex conjugate.

 \item We will also show in Sect. 5 that, in the {\it K\"ahler} case, the imaginary and real parts of the supercharge $Q$ in \p{Qcovgen}
can be interpreted as the Dirac operator $/\!\!\!\!{\cal D}$ and the operator $S$ defined in Eq. \p{DS}.

 \item Thus, for  K\"ahler manifolds, when the Lagrangians \p{genAct} and \p{N=1} coincide, two different ${\cal N} =2$
supersymmetry structures
$(/\!\!\!\!{\cal D}^2 \,;\  \ /\!\!\!\!{\cal D}\,,\  /\!\!\!\!{\cal D}  \gamma^{D+1})$ and
$(/\!\!\!\!{\cal D}^2 \,; \ \ /\!\!\!\!{\cal D}\,, \ S) $ are  possible.
Note that this does not imply an extended ${\cal N} =4$ supersymmetry because
the anticommutator $\{S, \  /\!\!\!\!{\cal D} \gamma^{D+1} \}$ does not vanish.

\item Note, however, that, for
{\it hyper-K\"ahler} manifolds where three different complex structures are present, one can construct three different new supercharges
 \be
S^{(f)} =  I^{M(f) }_{\ N} \gamma^N {\cal D}_M\ , \ \ \ \ \ \ \ \ \ \ f = 1,2,3\ ,
 \ee
such that the Lagrangian \p{LKahler},  \p{N=1} enjoys an ${\cal N} =4$ supersymmetry \break \cite{Wipf}
\footnote{This ${\cal N}=4$ supersymmetry holds also for more general class of sigma models associated with the so called
 HKT (hyper-K\"ahler with torsion) manifolds \cite{HKT}.
Their off-shell superfield description was given in a recent paper \cite{DelIv}.}.

\item There exists also an ${\cal N} = 4$  {\it completion} of the system \p{LKahler} for any K\"ahler manifold,
as will be discussed in Sect. 4.

\item For  non-K\"ahler manifolds, the sum ${\cal Q} = \bar Q + Q$ can also be interpreted
as a Dirac operator, but with some extra torsions. Indeed, one can show that
\be
\label{Qreal}
  {\cal Q}  \ =\ \sqrt{2} \psi^M  \left[ \Pi_M -  \frac i2\,  \Omega_{M, BC} \psi^B \psi^C
+ \frac i{12}\, C_{MKL}  \psi^K \psi^L   \right]
 \ ,
 \ee
which may be interpreted as a torsionful Dirac operator where the torsion tensor involves an
extra factor 1/3 compared to the Bismut connection \cite{Braden}.

  \end{itemize}

\subsection{Examples}
Here we consider two examples of SQM on complex manifolds.\\

\noindent{\bf 1. $\mathbb{CP}^n$ model.}
This is a  K\"ahler manifold,
so the torsion \p{Cikl}  vanishes
and many formulas look simpler. The corresponding K\"ahler potential is
\be
K = \ln (1 + z \bar z)\,, \quad z\bar z \equiv z{\,}^j\bar{z}{\,}^{\bar j}\,.
\ee
We choose
\be
W = - \frac {q}2 \, K = - \frac {q}2 \, \ln (1 + z \bar z)\,.
\ee
As we will see in Sect. 5, the quantum problem is consistent when the constant
$q$ is integer for odd $n$ and half-integer for even $n$.

The metric is given by the well known Fubini-Study expressions:
\bea
\lb{Fubini}
&& h_{j\bar k} = \partial_j \partial_{\bar k}\,\ln (1 + z \bar z)
= \frac{1}{1 + z\bar z}\left(\delta_{j\bar k} - \frac{z{\,}^k \bar z{\,}^{\bar j}}{1 + z\bar z} \right), \nn
&& h^{\bar k j} = (1 + z\bar z)\left(\delta^{\bar k j} + z{\,}^j\bar{z}{\,}^{\bar k} \right).
\eea
Note the  specific for  $\mathbb{CP}^n$ relation
\be
\lb{Kdet}
K \ =\ - \frac 1{n+1} \ln \det h\,.
 \ee
We choose the vielbeins in the form \cite{Wipf}:
\bea
&& e^a_l = \frac{1}{\sqrt{1 + z\bar z}}\left(\delta^a_l - \frac{z^a\bar z^{\bar l}}{\sqrt{1 + z\bar z}(1 + \sqrt{1 + z\bar z})} \right), \nn
&& e^l_b = \sqrt{1 + z\bar z}\left(\delta^l_b + \frac{z^l\bar z^{\bar b}}{1 + \sqrt{1 + z\bar z}} \right), \nn
&& e^{\bar l}_{\bar a} = \sqrt{1 + z\bar z}\left(\delta^{\bar l}_{\bar a} + \frac{z^a\bar z^{\bar l}}{1 + \sqrt{1 + z\bar z}} \right), \nn
&& e^{\bar a}_{\bar l} =  \frac{1}{\sqrt{1 + z\bar z}}\left(\delta^{\bar a}_{\bar l}
- \frac{z^l\bar z^{\bar a}}{\sqrt{1 + z\bar z}(1 + \sqrt{1 + z\bar z})} \right). \lb{CPn}
\eea

The supercharges \p{Qcovgen} in this special case look as follows
  \bea
Q^{\rm cov} = \sqrt{2}\psi^c e^k_c\left[{\tilde \Pi}_k
+ i \psi{\,}^b \bar\psi{\,}^{\bar a}  \omega_{k, \bar a b}
    \right], \quad
\bar Q^{\rm cov} = \sqrt{2}\bar\psi{\,}^{\bar c} e^{\bar k}_{\bar c}
\left[\bar {\tilde \Pi}_{\bar k}
+ i \bar\psi{\,}^{\bar a} \psi{\,}^{b} \bar{\omega}_{\bar k, b \bar a}
\right], \lb{Qcovgenspe}
\eea
where
\bea
&& \tilde \Pi_k = \frac{1}{i}\left[\frac{\partial}{\partial z^k}
+ \frac{1}{2}\left(q - \frac{n+1}{2}\right)\frac{\bar z{\,}^{\bar k}}{1 + z\bar z}\right], \nn
&& \hat{\bar\Pi}_{\bar k} = \frac{1}{i}\left[\frac{\partial}{\partial \bar z{\,}^{\bar k}}
- \frac{1}{2}\left(q + \frac{n+1}{2}\right)\frac{z^k}{1 + z\bar z}\right] \lb{PiCPn}
\eea
and $\omega_{k, \bar a b} \,, \  \bar\omega_{\bar k, b \bar a}$ were defined in Eq.(\ref{omkal}).

\vspace{3mm}

\noindent{\bf 2. $S^4$ model.} As a second example, we consider a 4-dimensional conformally
flat manifold  with the metric
 \be
\lb{metrS4}
ds^2 = \frac {2\, dz^j d\bar z^{\bar j}}{f^2}\, ,\ \ \ \ \ \ \ \ \ \ \ \ \ \ j =1,2 \ .
 \ee
When $f = 1 + z \bar z$, this is the metric of $S^4$ or rather
$S^4\backslash \{\cdot\}$ (the metric (\ref{metrS4}) being singular in infinity).
Under a natural choice of vielbeins,  $\det e = \det \bar e = 1/f^2$ and the non-zero components of the spin connection
$\Omega_{k, \bar a b}$ are
\be
\lb{connS4}
 \Omega_{1, \bar 1 1} =  \Omega_{2, \bar 2 1} \ =\ - \partial_1 \ln f \, , \ \ \ \ \ \ \
 \Omega_{1, \bar 1 2} =  \Omega_{2, \bar 2 2} \ =\ - \partial_2 \ln f \ .
 \ee
This is not a K\"ahler manifold. Taking the general expression (\ref{Qcovgen}) for the supercharges, we derive
for $W = 0\,$,
 \be
\lb{QS4}
{\cal Q} = -i\sqrt{2} \psi^a [ f \partial_a - (\partial_a f)] - i\sqrt{2} \psi^1 \psi^2[ (\partial_2 f) \bar \psi^1 -
(\partial_1 f) \bar \psi^2 ]\, .
 \ee

An ${\cal N} = 4$ SQM model describing the motion over any conformally flat 4-dimensional manifold with the metric (\ref{metrS4})
 with or without background gauge field
was constructed in \cite{Konush} based on the action given in \cite{IO} (see also \cite{IKS}).
In the case when the gauge field is absent, the {\it flat} (in the Hilbert space sense, as discussed above)
supercharges have the form
  \bea
\lb{Q-KS}
Q_\alpha = f(\sigma_\mu \bar \psi)_\alpha P_\mu   - i (\partial_\mu f)  \psi_\gamma \bar\psi^\gamma (\sigma_\mu \bar \psi)^\alpha\,,
\nonumber \\
\bar Q^\alpha = (\psi \sigma_\mu^\dagger)^\alpha P_\mu f + i (\partial_\mu f)
(\psi \sigma_\mu^\dagger)^\alpha \psi_\gamma \bar\psi^\gamma \, ,
 \eea
where $\sigma_\mu = (i, \vec{\sigma}) $, \ $\sigma_\mu^\dagger = (-i, \vec{\sigma})$.

It is straightforward to see that,
after performing the similarity transformation (\ref{similar}),
 the supercharge (\ref{QS4}) coincides with $\bar Q^1$ in (\ref{Q-KS})
 under the identification
$$ z^1 = \frac {x_3 + ix_4}{\sqrt{2}}\, , \ \ \ \ \ \ \ \ \    z^2 = \frac {x_1 - ix_2}{\sqrt{2}}\, , $$
or with $\bar Q^2$, under the identification
$$ z^1 = \frac {x_1 + ix_2}{\sqrt{2}}\, , \ \ \ \ \ \ \ \ \    z^2 = \frac {ix_4 - x_3}{\sqrt{2}}\ .$$
These two possibilities reflect the presence of two different ${\cal N}=2$ Poincar\'e superalgebras in the ${\cal N}=4$ superalgebra.

In the case of  $S^4\backslash \{\cdot\}$, the spectrum of the model was recently analyzed in \cite{S4}. In spite of the singularity,
which could ruin the supersymmetry \cite{flux}, this does not happen in this case. The spectrum is supersymmetric, involving
3 bosonic zero modes.

\section{Completion to K\"ahler ${\cal N}=4$ SQM model}
\setcounter{equation}0

Our starting point is the ${\cal N}=2$ SQM model with the superfield Lagrangian
${\cal L}_\sigma$ in \p{start} involving the K\"ahler metric \p{Kahlmetr}.
We do not add the gauge part ${\cal L}_{gauge}$. So we choose
\be
h_{j\bar k}(Z, \bar Z) = \partial_j\partial_{\bar k}\,K(Z, \bar Z)\,, \quad  W = 0
\ee
in \p{start}. The corresponding component Lagrangian was written in \p{LKahler}.

Using the chirality properties of $Z^j, \bar Z{\,}^{\bar k}$ and the algebra of ${\cal N}=2$ spinor derivatives, it will be convenient
to rewrite the corresponding superfield Lagrangian in the following three equivalent (they coincide up to a total time derivative) forms:
\bea
{\cal L}^{K} &=& -\frac{1}{4}\partial_j\partial_{\bar k}\,K(Z, \bar Z)\,D Z^j\, \bar D\bar Z{\,}^{\bar k} \simeq
- \frac{i}{2}\,\dot{Z}{\,}^j\partial_j\, K \nn
&\simeq& \frac{i}{2}\,\dot{\bar Z}{\,}^{\bar k}\partial_{\bar k}\, K  \simeq \frac{i}{4}\left(\dot{\bar Z}{\,}^{\bar k}\partial_{\bar k}\, K
-\dot{Z}{\,}^j \partial_j\, K\right). \lb{equiv}
\eea

Now consider an extended Lagrangian
\be
\tilde{{\cal L}}^K = {\cal L}^{K} + \frac{1}{4}\,h_{j\bar k}\,\Phi^j\,\bar\Phi^{\bar k}\,, \lb{exten}
\ee
where $\Phi^j\,, \ \bar\Phi^{\bar k}$
are chiral and anti-chiral fermionic ${\cal N}=2$ (0+1)-dimensional superfields, $\bar D\,\Phi^j = D\,\bar\Phi^{\bar k} = 0\,$.
It is straightforward to check that \p{exten} is
invariant, modulo a total derivative, under the following
extra ${\cal N}=2$ supersymmetry transformations:
\be
\delta Z^j = - \zeta\, \Phi^j\,, \quad \delta \bar{Z}^{\bar k} = \bar\zeta\, \bar\Phi{\,}^{\bar k}\,, \quad
\delta \Phi^j = 2i\,\bar\zeta\, \dot{Z}{\,}^j\,, \quad \delta \bar\Phi{\,}^{\bar k} = - 2i\,\zeta\, \dot{\bar Z}{\,}^{\bar k}\,. \lb{extrasusy}
\ee
 These variations form the same algebra with respect to Lie brackets as the variations \p{transfcomp}
corresponding to the manifest
world-line ${\cal N}=2$ supersymmetry. Thus, they extend the latter to off-shell (0+1)-dimensional  ${\cal N}=4$ supersymmetry.

The superfields $\Phi^j\,, \bar\Phi^{\bar k}$ have the following $\theta$ expansions
\be
\Phi^j = \sqrt{2}\chi{\,}^j + \theta d^j - i\sqrt{2}\theta\bar\theta\,\dot{\chi}{\,}^j\,, \quad \bar\Phi^{\bar k} =
\sqrt{2}\bar\chi{\,}^{\bar k} + \bar\theta \bar{d}{\,}^{\bar k} + i\sqrt{2}\theta\bar\theta\,\dot{\bar\chi}{\,}^{\bar k}\,. \lb{expPhi}
\ee
We observe  that they contain no new bosonic fields of physical dimension, only the auxiliary bosonic fields
$d{\,}^j, \bar{d}{\,}^{\bar k}$ as well as the extra physical fermionic fields $\chi{\,}^j, \bar\chi{\,}^{\bar k}$.
Thus, in this model we deal with $n$ off-shell
${\cal N}=4$ supermultiplets $({\bf 2, 4, 2})$, the subsequent numerals standing, respectively, for the numbers of the physical bosonic,
physical fermionic and auxiliary bosonic fields\footnote{In this notation, the ${\cal N}=2$ multiplets corresponding
to  the superfields $Z^i$ and $\Phi^i$
can be denoted as ${\bf (2, 2, 0)}$ and ${\bf (0, 2, 2)}$.}. The manifest ${\cal N}=2$ supersymmetry acts on the component fields in \p{expPhi} as
\be
\delta\chi^j = -\frac{1}{\sqrt{2}}\epsilon\,d^j\,, \quad \delta d^j = 2\sqrt{2}i\bar\epsilon\,\dot{\chi}{\,}^j\,, \quad
\delta\bar{\chi}^{\bar k} = -\frac{1}{\sqrt{2}}\bar\epsilon\,\bar{d}{\,}^{\bar k}\,, \quad
\delta \bar{d}{\,}^{\bar k} = 2\sqrt{2}i\epsilon\,\dot{\bar\chi}{\,}^{\bar k}\,. \lb{mansusy2}
\ee
The second supersymmetry transformations \p{extrasusy} has the following realization in components:
\bea
&& \delta z{\,}^j = -\sqrt{2}\,\zeta\,\chi{\,}^j\,, \; \delta\psi{\,}^j = \frac{1}{\sqrt{2}}\zeta\,d{\,}^{j}\,, \;
\delta \chi{\,}^j = \sqrt{2}i\bar\zeta\,\dot{z}{\,}^j\,, \;
\delta d{\, }^j = -2\sqrt{2}i\bar\zeta\,\dot{\psi}{\,}^j\,, \nn
&& \delta \bar{z}{\,}^{\bar k} = \sqrt{2}\bar\zeta\,\bar{\chi}{\,}^{\bar k}\,, \; \delta\bar\psi{\,}^{\bar k}
= \frac{1}{\sqrt{2}}\bar\zeta\,\bar{d}{\,}^{\bar k}\,, \;
\delta \bar\chi{\,}^{\bar k} = -\sqrt{2}i\zeta\,\dot{\bar z}{\,}^{\bar k}\,, \;
\delta \bar{d}{\,}^{\bar k} = -2\sqrt{2}i\zeta\,\dot{\bar\psi}{\,}^{\bar k}\,. \lb{extracomp}
\eea

After going to the component fields in the action corresponding to the modified superfield Lagrangian \p{exten} and eliminating
the auxiliary fields $d{\,}^j, \bar{d}{\,}^{\bar k}$ by their equations of motion,
\be
d{\,}^j = 2h^{\bar p j}\partial_l\,h_{t\bar p}\,\chi{\,}^t\psi{\,}^l\,, \quad
\bar{d}{\,}^{\bar k} = 2h^{\bar k p}\partial_{\bar l}\,h_{p\bar j}\,\bar\psi{\,}^{\bar l}\bar\chi{\,}^{\bar j}\,, \lb{auxd}
\ee
the contribution of the second term in \p{exten} to the total component Lagrangian reads:
\be
\Delta L = \frac{i}{2}\,h_{j\bar k}\left(\chi{\,}^j \nabla{\bar \chi}{\,}^{\bar k} -
\nabla{\chi}{\,}^j \bar \chi{\,}^{\bar k}\right)
+ R_{j\,\bar k\, l\,\bar p}\,\psi{\,}^j \bar\psi{\,}^{\bar k} \chi{\,}^l\bar\chi{\,}^{\bar p} \,. \lb{actadd}
\ee
Here,
\be
\nabla{\bar \chi}{\,}^{\bar k} = \dot{\bar \chi}{\,}^{\bar k} + \dot{\bar z}{\,}^{\bar p}\Gamma^{\bar k}_{\bar p\, \bar j}{\bar \chi}{\,}^{\bar j}\,, \quad
 \nabla{\chi}{\,}^j = \dot{\chi}{\,}^j + \dot{z}{\,}^{l}\Gamma^{j}_{l\, p}{\chi}{\,}^{p}\,,
\ee
and  $R_{j\,\bar k\, l\,\bar p}$ is the Riemann tensor for the K\"ahler metric defined in (\ref{Riemann}). Its appearance in the Lagrangian
is an important new feature of the ${\cal N} = 4$ case compared to Eq.\p{LKahler}.

The total ${\cal N}=4$ supersymmetric component Lagrangian can be concisely written as
\be
L = \ h_{j\bar k}\left[ \dot{z}{\,}^j\dot{\bar z}{\,}^{\bar k} +
\frac{i}{2} \left( \psi{\,}^j\nabla{\bar \psi}{\,}^{\bar k} +
\chi{\,}^j \nabla{\bar \chi}{\,}^{\bar k}
- \nabla{\psi}{\,}^j \bar \psi{\,}^{\bar k}- \nabla{\chi}{\,}^j \bar \chi{\,}^{\bar k}\right) \right]\
+ R_{j\,\bar k\, l\,\bar p}\,\psi{\,}^j \bar\psi{\,}^{\bar k} \chi{\,}^l\bar\chi{\,}^{\bar p}\,. \lb{acttot}
\ee
The ${\cal N}=4$ supersymmetry closes on shell, since we have eliminated the auxiliary
fields $d{\,}^j, \bar{d}{\,}^{\bar k}$.

 The Lagrangian \p{acttot} is well known. It coincides with the Lagrangian obtained by deleting spatial derivatives in
the (1+1)-dimensional ${\cal N} = 2$ $\sigma$-model Lagrangian \cite{Zumino} and discussed, e.g., in \cite{Davis,Macfar,howto}
 (there, fermionic fields $\psi{\,}^j, \chi{\,}^j$ were combined into a SU(2) doublet).
We refer the reader to \cite{howto}
 for the expressions for the classical and quantum supercharges, the Hamiltonian, etc.

It is worth also recalling
 that the Lagrangian \p{acttot} coincides with the generic SQM sigma-model Lagrangian involving $D$ supermultiplets ({\bf 1}, {\bf 2}, {\bf 1} )
\cite{sigma1,sigma2,sigma3},
 \be
\label{sigmareal}
 L \ =\ g_{MN} \left( \frac 12  \dot z^M \dot z^N  + i\, \bar \psi^M \nabla \psi^M \right) + \frac 12
R_{MNPQ} \, \bar \psi^M \psi^N \bar \psi^P \psi^Q \ .
 \ee
  For a generic metric, the latter Lagrangian enjoys only ${\cal N} = 2$ supersymmetry, but
in the K\"ahler case, a second pair of supercharges can be found. Note also that, when an external gauge field is present,
there is no such second pair. A related almost equivalent statement is that
no ${\cal N}=4$ completion based on the linear chiral ${\cal N}=4, d=1$ multiplets ${\bf (2, 4, 2)}$ is possible
for the theory (\ref{start}) with $W \neq 0\,$. Note that such a completion becomes possible, if extending the ${\cal N}=2$
chiral multiplets ${\bf (2, 2, 0)}$ to {\it nonlinear} versions of the ${\cal N}=4$ multiplets ${\bf (2,4,2)}$ or
${\bf (4,4,0)}$ \cite{nonl1,nonl2,DelIv}.

\section{Quantum supercharges and geometry}
\setcounter{equation}0

Let us  assume that $\det \bar e$ = $\det e$ = $\sqrt{\det \, h}\;$\footnote{Such a choice amounts to fixing a gauge with respect
to the local frame U(1) transformations of the vielbeins.} and
choose
 \be
\lb{choiceW}
W = \frac {q}{2 (n+1)} \, \ln \det h\, .
 \ee
 Then the general supercharges \p{Qcovgen} take  the form
(\ref{Qcovgenspe}), (\ref{PiCPn}) where we should replace $\omega_{k, \bar a b} \to \Omega_{k, \bar a b}$ and
 \bea
\lb{replace}
\frac {\bar z^k}{1+ z \bar z} \ \rightarrow \  - \frac 1{n+1} \partial_k \left( \ln \det h \right), \nonumber \\
 \frac  {z^k}{1+ z \bar z} \ \rightarrow \  - \frac 1{n+1} \partial_{\bar k} \left( \ln \det h \right).
 \eea
We see that there are special values $q = \pm (n+1)/2$ where either ${\tilde \Pi}_k$ or $\bar {\tilde \Pi}_k$ coincide with
the usual holomorphic or antiholomorphic derivatives. Consider first the case $q = (n+1)/2$. It is not difficult to check
that the action of $Q^{\rm cov}$ on the wave functions
 \be
\lb{wave}
\Psi(z^k, \bar z^k; \psi^a) \ =\ A^{(0)}(z^k, \bar z^k) + \psi^a A^{(1)}_a(z^k, \bar z^k) + \ldots
+ \psi^{a_1}\cdots \psi^{a_n}A^{(n)}_{[a_1 \cdots a_n]}(z^k, \bar z^k)
 \ee
is isomorphic to the action of the exterior holomorphic derivative $\partial$ on the set of $n+1$ holomorphic (p,0)-forms (the
term $\propto \Omega$ in  $Q^{\rm cov}$ cancels out the term coming from differentiation of the vielbeins in virtue
of the structure equation \p{Cartan1}).
The Hermitian-conjugate operator $\bar Q^{\rm cov}$ is then isomorphic to
$\partial^\dagger$. In other words, in this case the supercharges (\ref{Qcovgenspe}) realize the standard untwisted (i.e. involving
no additional gauge field) Dolbeault complex.

Likewise, in the case $q = -(n+1)/2$, the action of the operator $\bar Q^{\rm cov}$ on {\it anti}-holomorphic wave
functions $\Psi(z^k, \bar z^k; \bar\psi^a)$ is isomorphic to the action of the operator $\bar\partial$ on antiholomorphic
(0,p)-forms, the operator $Q^{\rm cov}$ playing the role of $\bar\partial^\dagger$. Thus, in this case we are dealing with
the anti-holomorphic untwisted Dolbeault complex.

For any other value of $q$, an extra Abelian gauge field is present in the framework of both the holomorphic
and antiholomorphic Dolbeault interpretations, i.e.
\be
A_k = \frac{i}{4}\left(1 - \frac{2q}{n +1}\right)\partial_k \ln \det h  \lb{twistDol}
\ee
in the holomorphic case and
 \be
A_{\bar k} = \frac{i}{4}\left(1 + \frac{2q}{n +1}\right)\partial_{\bar k} \ln \det h \lb{twistDolbar}
\ee
 in the antiholomorphic case.
We face what is called twisted Dolbeault complex.

Until now we dealt with the generic (non-K\"ahler) ${\cal N}=2$ SQM model, the only restriction was the relation \p{choiceW}.
If the manifold is K\"ahler, the supercharges admit {\it another} even more interesting geometric interpretation:
 when $q=0$, the {\it sum}
$Q^{\rm cov} + \bar Q^{\rm cov}$ can be interpreted as the untwisted Dirac operator. When $q \neq 0$, an extra
Abelian gauge field is present.

Indeed, the standard untwisted Dirac operator in the real notations is [cf. \p{Qreal1}]
  \be
\lb{realDirac}
/\!\!\!\!{\cal D} \ =\ \gamma^A e^M_A \left( \partial_M + \frac 14 \Omega_{M, BC}
\gamma^B \gamma^C \right) \equiv \gamma^M{\cal D}_M\,.
 \ee
When splitting $M \equiv (k, \bar k),\ \ A \equiv (a, \bar a)$ and introducing $\sqrt{2}\psi^a \equiv \gamma^a,\
\sqrt{2}\bar\psi^{\bar a} \equiv \gamma^{\bar a}$, one can be convinced
that, for K\"ahler manifolds, one can represent
 \be
\label{decomp}
 /\!\!\!\!{\cal D} \ =\ /\!\!\!\!{\cal D}^{\rm Hol} - \left(  /\!\!\!\!{\cal D}^{\rm Hol} \right)^\dagger \ ,
 \ee
 where
 \be
\lb{DirHol}
 /\!\!\!\!{\cal D}^{\rm Hol} \ =\ \sqrt{2}\psi^b e^k_b \left[\partial_k + \frac 12 \omega_{k, \bar a d} (\bar\psi{\,}^{\bar a}\psi{\,}^d -
 \psi{\,}^d \bar\psi{\,}^{\bar a})\right]
 \ee
and
\be
\lb{DirHol2}
 \left( /\!\!\!\!{\cal D}^{\rm Hol}\right)^\dagger  \ =\ -\sqrt{2}\bar\psi^{\bar b} e^{\bar k}_{\bar b} \left[\partial_{\bar k} +
 \frac 12 \bar \omega_{\bar k, a \bar d} (\psi{\,}^{a}\bar\psi{\,}^{\bar d} -
 \bar\psi{\,}^{\bar d} \psi{\,}^{a})\right].
 \ee
These operators coincide, up to the factor $i$, with the supercharges (\ref{Qcovgen}), \p{PidW} in which one chooses $\Omega = \omega$
and $W = 0$:
\be
 /\!\!\!\!{\cal D}^{\rm Hol} \ =\ i Q^{cov}\,, \quad \left( /\!\!\!\!{\cal D}^{\rm Hol}\right)^\dagger = -i \bar Q{\,}^{cov}\,, \qquad ( C_{ki\bar l} =q  =0)\,.
\ee
For $q \neq 0$, an additional Abelian gauge field is present, and we are facing the twisted Dirac operator in this case. Note
that the definition of ``twisting'' or ``untwisting'' is different in the interpretations in terms of Dolbeault and Dirac complexes.
E.g., the choice $q=0$ corresponds to an {\it untwisted} Dirac complex, but to the {\it twisted} Dolbeault complex (as is seen from \p{twistDol}, \p{twistDolbar}).

The operator $ /\!\!\!\!{\cal D}$ is anti-Hermitian. Consider now the {\it real} part of $/\!\!\!\!{\cal D}^{\rm Hol}$,
\be
S \ =\   /\!\!\!\!{\cal D}^{\rm Hol} + \left( /\!\!\!\!{\cal D}^{\rm Hol} \right)^\dagger.
 \ee
One can be convinced that, instead of $\sqrt{2}\psi^M{\cal D}_M \equiv  \gamma^M{\cal D}_M$
for the imaginary (anti-Hermitian) part of $/\!\!\!\!{\cal D}^{\rm Hol}\,$, for the real (Hermitian) part we obtain
the expression
 \be
\label{S}
S = \ -i \gamma^N I^M_{\ N} {\cal D}_M  \ .
 \ee
This is immediately seen when writing the components of the complex structure tensor \p{comstruc} in the complex
basis:
$I^m_{\ n} = i\delta^m_n; \
I^{\bar m}_{\ \bar n} = -i\delta^{\bar m}_{\bar n}$.
 The pair of supercharges \p{DS} is thus reproduced.

A by-product of this analysis is a physical proof of the purely mathematical fact:
for K\"ahler manifolds, the twisted Dirac complex is equivalent
to the twisted Dolbeault complex, bearing in mind that the twisting (the adding of Abelian gauge fields)
in the Dirac complex and in the Dolbeault
complex is different. This fact is known to mathematicians, see e.g.
the Propositions 1.4.23 and 1.4.25 in the book \cite{Nicola}.

If the manifold is not K\"ahler, the decomposition (\ref{decomp}) - (\ref{DirHol2}) is no longer valid.
However, as was noted above, the sum $\bar Q + Q$ of the quantum supercharges \p{Qcovgen} can be represented as
 the Dirac operator
with some special torsions \p{Qreal}.
The imaginary part $S \propto  \bar Q -  Q$ may be obtained from
\p{Qreal}
 by commuting \p{Qreal}  with the fermion charge operator
$F = \frac i2 I_{MN}\psi^M \psi^N  =  \frac i4 I_{MN}\gamma^M \gamma^N  $.
 The result is \cite{SmHKT}
 \be
\lb{Sgen}
S  \ =\ \sqrt{2} \psi^N I^M_{\ N}  \left[ \Pi_M -  \frac i2\,  \Omega_{M, BC} \psi^B \psi^C
- \frac i{4}\, C_{MKL}  \psi^K \psi^L   \right] \, .
 \ee

\section{Index}
\setcounter{equation}0

The Euclidean path integral representation for the index (\ref{indWit}) of our system is
 \bea
\lb{indpathH}
I &=& \int \prod_{j\tau} \frac {d\pi_j(\tau) d\bar\pi_{\bar j}(\tau) dz^j(\tau)   d\bar z^{\bar j}(\tau)}{(2\pi)^2} \prod_{a\tau}
d \psi^{ a} (\tau)  d \bar\psi^{\bar a} (\tau) \nonumber \\
&& \times \,\exp \left\{ \int_0^\beta d\tau \left[ i \pi_j \dot z^j +
i \bar \pi_{\bar j} \dot {\bar z}^{\bar j} +  \dot{\bar \psi}^{\bar a} \psi^a  - H(\pi_j,
\bar \pi_{\bar j}, z^j, \bar z^{\bar j};  \bar\psi^{\bar a}, \psi^a) \right ] \right\} \ ,
 \eea
where both bosonic {\it and} fermionic variables satisfy the periodic boundary conditions,
$z^j(\beta) = z^j(0)$, etc. Expand all the variables  in the Fourier series,
 \be
\lb{Fourier}
z^j(\tau) =  z^{j(0)} + \sum_{m \neq 0} z^{j(m)} e^{2\pi i m \tau/\beta} \ \ \ \ \ \
\bar z^{\bar j}(\tau) =  z^{\bar j(0)} + \sum_{m \neq 0} \bar z^{\bar j(m)} e^{-2\pi i m \tau/\beta} \ ,
 \ee
and similarly for $\pi_j(\tau), \bar \pi_{\bar j}(\tau)$ and $\psi^a(\tau), \bar\psi^{\bar a}(\tau)$.
If $\beta$ is small, we  {\it seemingly} (see below) can neglect the
nonzero modes in the expansion, neglect thereby the terms with time derivatives in (\ref{indpathH}), and rewrite
(\ref{indpathH}) as an {\it ordinary} integral \cite{Cecotti,Mukhi}:
 \bea
\lb{ordinary}
I &=& \int \prod_j \frac {d\pi_j^{(0)} d\bar \pi_{\bar j}^{(0)} dz^{j(0)}  d\bar z^{j(0)}}{(2\pi)^2} \prod_a  d\psi^{a(0)} d\bar \psi^{\bar a(0)}
\nn
&&\times \,\exp \left \{  - \beta H(\pi_j^{(0)}, \bar \pi_{\bar j}^{(0)}, z^{j(0)},  \bar z^{\bar j(0)} ;
\psi^{a(0)}, \bar\psi^{\bar a (0)} ) \right \}.
 \eea
The functional integral is reduced to the ordinary one in the semiclassical limit $\beta \to 0$. However, the index
(\ref{indWit}) does not depend on $\beta$, and the estimate (\ref{ordinary}) should be true for any $\beta$.

Substituting here the Hamiltonian (\ref{Hclass}) with the choice (\ref{choiceW}) (remember that, for K\"ahler manifolds,
the last term in (\ref{Hclass}) vanishes), we can easily integrate over $\prod_{j} d\pi_j d\bar \pi_{\bar j}$ and over
$\prod_{a} d\psi^a d\bar \psi^{\bar a}  $ to obtain
 \be
\lb{indF}
 I \ =\ \left( \frac {1}{2\pi} \right)^n \int  \prod_j dz^j d\bar z^{\bar j}  \det \| h_{j \bar k} \| \det \|i{\cal F}_{a \bar b} \| \ ,
 \ee
where $i{\cal F}_{a \bar b} =  -2e^j_a e^{\bar k}_{\bar b} \partial_j \partial_{\bar k} W $ is related to the 2-form describing the
magnetic field strength. In the simplest  $\mathbb{CP}^n$ case under the choice (\ref{choiceW}), we have $i{\cal F}_{a \bar b} =
q \delta_{a \bar b}$ leading to
 \be
  I \ =\ \left( \frac {q}{2\pi} \right)^n \int  \, \prod_j dz^j d\bar z^{\bar j} \,   (\det \, h)\, .
 \ee
The calculation with the Fubini-Study metric (\ref{Fubini}), i.e. with $\det h = \frac{1}{(1 + z\bar z)^{n+1}}\,$, gives
 \be
\lb{indnaive}
I_{\mathbb{CP}^n} \ \stackrel ? = \ \frac {q^n}{n!} \ .
 \ee
This result looks suspicious. Indeed, to make it integer (the index should be integer for the Dirac
operator to make sense: only in this case the manifold admits {\it spin structure}), $q$ should depend on $n$ in an odd way.

 Actually, the estimate (\ref{indnaive}) is {\it wrong}. The correct estimate reads
\cite{AS1,AS2,AS3,AS4,Alv,FW}
 \be
\lb{indAS}
I \ =\ \int e^{{\cal F}/2\pi} {\det}^{-1/2} \left[ \frac {\sin \frac {\cal R} {4\pi}}  { \frac {\cal R} {4\pi}} \right],
 \ee
where ${\cal F}$ is the field strength 2-form and ${\cal R}$ is the matrix
2-form associated with the Riemann curvature,
   \be
\lb{FR}
 {\cal F} = \frac 12 F_{MN}\, dx^M \wedge dx^N \ , \ \ \ \ \ {\cal R}^{AB} = \frac 12 R^{AB}_{\ \ \ MN}\, dx^M \wedge dx^N \ ,
 \ee
$A,B$ being the tangent space indices.

The precise meaning of the representation \p{indAS} is that the volume integral
in its r.h.s. projects out only the forms of the maximal  rank $D$ from the Taylor
expansion of the integrand.  Thus, for 4-dimensional manifolds, the index is represented as the sum of two terms,
  \be
\lb{ind4}
 I_{d=4} \ =\ \frac 1{8\pi^2}  \int {\cal F} \wedge   {\cal F}  + \frac 1{192\pi^2} \int  {\rm Tr} \{ {\cal R} \wedge {\cal R} \}\,.
 \ee
The topological invariants in the r.h.s. are known as the second Chern class $c_2$ and the Hirzebruch signature $\tau$
(the latter enters with the coefficient $-1/8$).
For higher dimensions, the index is a sum of many different invariants.

It is convenient to represent the determinant factor in (\ref{indAS}) as
 \be
\lb{detlam}
{\det}^{-1/2}[\cdots] \ =\ \prod_{\alpha=1}^{n} \frac {\lambda_\alpha/(4\pi)}{ \sinh(\lambda_\alpha/(4\pi)} \ ,
 \ee
where $\lambda_\alpha$ are the eigenvalues of the antisymmetric matrix ${\cal R}^{AB}$. This can be derived by diagonalizing,
 $$
{\cal R} \ \longrightarrow (i\sigma_2 \lambda_1, \ldots, i\sigma_2 \lambda_n)\, ,
 $$
and noting that, for any {\it even} function $f({\cal R})$,
 $$
{\det}^{-1/2} f({\cal R}) \ =\  \prod_{\alpha=1}^{n} \frac 1 {f(i\lambda_\alpha)} \ .
 $$

The estimate (\ref{indnaive}) for  $\mathbb{CP}^n$ would be reproduced, if ignoring this   curvature-dependent
determinant factor
in (\ref{indAS}). When including this factor, we obtain instead
 \be
\lb{ICPn}
I_{\mathbb{CP}^n} \ =\ \left( \begin{array}{c} q + (n-1)/2 \\ n \end{array} \right),
 \ee
where $q$ must be integer for odd $n$ and half-integer for even $n$. The index is given by Eq.(\ref{ICPn}) when
$q \geq (n+1)/2$. For negative $q \leq -(n+1)/2$,
it is given by
 \be
\lb{ICPnneg}
 I_{ \mathbb{CP}^n}(q < 0)  \ =\ (-1)^n \left( \begin{array}{c} |q| + (n-1)/2 \\ n \end{array} \right).
 \ee
  The index vanishes for $|q| < (n+1)/2$ \footnote{Note in passing that the index (\ref{ICPn}) is closely related
  to the Witten index in 3d supersymmetric Yang-Mills-Chern-Simons theory
\cite{WitSYMCS}. See \cite{SmSYMCS1,SmSYMCS2} for detailed discussion.}.

The result \p{ICPn} for the index in  $\mathbb{CP}^n$ can also be derived  directly,
 simply by counting   the number of independent ground states \cite{imt,Wipf}, i.e. the number of the normalized
(with the measure \p{normcov}, in which $\det h = 1/(1+z \bar z)^{n+1}$ ) solutions to the equations
\be
Q^{cov} \Psi_0 = \bar{Q}{}^{cov}\Psi_0 = 0\,, \lb{cpGr}
\ee
with $Q^{cov}$ and $\bar Q^{cov}$ defined in \p{Qcovgenspe}, \p{PiCPn}.
   Choosing, e.g., the holomorphic representation \p{wave} for the wave functions, we find that, in the sector
of {\it zero} fermionic charge, the equation
$\bar{Q}{}^{cov}\Psi_0 = 0$ is satisfied identically, while the equation $Q^{cov}\Psi_0 = 0$ implies
 \be
\partial_k \Psi_0 = - \frac {s \bar z^{\bar k}}{1 + z \bar z} \Psi_0\,,
 \ee
with
\be
2s = q - \frac{n+ 1}{2}\,.
\ee
We see that the normalized solutions exist only at $s \geq 0$. They have the form
\be
\Psi_0 = \Psi(z, \bar z) = ( 1 + z \bar z)^{-s}\Phi(\bar z)\,,
\ee
where
 $\Phi(\bar z)$ is a polynomial of $\bar z^{\bar j}$
of the rank not higher than $2s$. Then the number of independent ground states is given by the binomial coefficient
\be
\frac{(n + 2s)!}{n!(2s)!} = \frac{(q + \frac{n-1}{2})!}{n!(q - \frac{n+ 1}{2})!}\, ,
\ee
which exactly coincides with Eq. \p{ICPn}. For negative $q$, the vacuum states are present in the sector of fermion
charge $F=n$, hence the factor $(-1)^n$ in Eq. \p{ICPnneg}.

What was wrong then in the calculation having led to (\ref{indnaive})?
The answer is that the  recipe \cite{Cecotti,Mukhi} that allowed us to replace the functional integral
(\ref{indpathH}) by the ordinary integral (\ref{ordinary}) and that works well for many SQM and supersymmetric field
theory systems {\it fails} in this case. To obtain the correct estimate for the index, one should take into account
the nonzero Fourier modes in the expansion (\ref{Fourier}) and integrate them over in the {\it Gaussian} (see below)
  approximation. This integral gives exactly the determinant factor in (\ref{indAS}).

To perform the actual calculation
\footnote{It is rather
similar in spirit to the calculation of the functional integral for SQM describing
the complex $(/\!\!\!\!{\cal D}^2 ;\    /\!\!\!\!{\cal D},  /\!\!\!\!{\cal D} \gamma^5)$ \cite{Alv,FW,Wind}.
We do it, however, in a much more detailed way.},
we assume $\beta$ to be small, impose periodic boundary conditions,  subdivide the interval $(0, \beta)$ into a large
number $N$ of integration points and integrate first over $\prod_{j\tau} \frac{d\pi_j(\tau)d\bar \pi_{\bar j}(\tau)}{2\pi}$ to obtain
 \bea
\lb{poslepi}
I &=&  \int \prod_\tau  \det h(\bar z^{\bar j}(\tau), z^j(\tau)) \prod_{j} \frac {d\bar z^{\bar j}(\tau) dz^j(\tau)}{2\pi (\beta/N) }
 \prod_{a} d\psi^{ a}(\tau) d\bar\psi^{\bar a}(\tau) \nn
&& \times \,\exp \left\{ - \int_0^\beta L_E(\tau) d\tau \right \},
 \eea
with
\bea
\lb{LE}
L_E = h_{j\bar k} \dot{z}^j \dot{\bar z}^{\bar k} + \frac 12 (\psi^a \dot{\bar\psi}^{\bar a} - \dot{\psi}^a \bar\psi^a )
+ \left(\dot{\bar z}^{\bar k} \omega_{\bar k, a \bar b}- \dot{z}^k \omega_{k,\bar b a} \right)  \psi^a \bar \psi^{\bar b} \,.
 \eea
 The product $\prod_\tau$ in \p{poslepi} runs over $N$ discrete points $\tau_r = \beta r/N, \ \ \ r = 0,\ldots, N-1\,$. For simplicity, we suppressed
the gauge part that was already successfully handled earlier by the Cecotti-Girardello method. It is the determinant factor depending
only on the Riemannian manifold geometry that is of interest for us now.

Substitute now the expansion (\ref{Fourier}) into (\ref{LE}). If the number of points $N$ is large   but finite,
we have also to keep the number of Fourier modes finite, so that the sum in
\p{Fourier} runs over $m = -M, \ldots, 0, \ldots, M$, where $N = 2M+1$. To calculate the functional
integral in the Gaussian approximation
\footnote{Incidentally, the result \p{poslepi} can also be reproduced by trading the
variables $\pi_j(\tau_r)$ and $\bar \pi_{\bar j} (\tau_r)$  for their Fourier
modes and performing  then the Gaussian integral over $\prod_m' d\pi_j^{(m)} d\bar \pi^{(m)}_{\bar j} $.
In this case, the factor $N^N$ seen in \p{poslepi} appears as the Jacobian of the variable change \p{Fourier}.}
(we will justify the validity of this approximation later),
we keep only quadratic (in $\bar z^{\bar j}_m, z^j_m, \bar \psi^{\bar a}_m$ and
$\psi^a_m$) terms and do the $\tau$-integral. The quadratic part of the Lagrangian gives
  \bea
 \lb{LE2}
\int^\beta_0  L_E^{(2)} d\tau \ &=&\ -i \beta \sum'_m \Omega_m \psi^a_m \bar\psi^{\bar a}_m -i \beta \sum'_m \Omega_m \left[
\psi^{a (0)} \omega_{k, \bar b a }^{(0)} \bar \psi_m^{\bar b} z^k_m - \bar \psi^{\bar b (0)} \psi_m^a \bar z_m^{\bar k}
\omega_{\bar k, a \bar b}^{(0)} \right] \nonumber \\
 && +\, \beta \sum'_m z^j_m \bar z^{\bar k}_m \left[ \Omega_m^2 h_{j \bar k}^{(0)} - i\Omega_m \psi^{a(0)} \bar \psi ^{\bar b (0)}
(\partial_j \omega_{\bar k, a \bar b}^{(0)} + \partial_{\bar k} \omega^{(0)}_{j, \bar b a} )  \right] \nonumber \\
&& -\, i\beta \sum'_m \Omega_m \left[\psi^{a (0)} \omega_{\bar k, a \bar b}^{(0)} \bar z^{\bar k}_m \bar \psi_{-m}^{\bar b}
- \bar \psi^{\bar a (0)} z^k_m \psi_{-m}^b \omega_{k, \bar a b}^{(0)} \right] \nonumber \\
&& -\, i\beta \sum'_m \Omega_m\,\psi^{a (0)}\bar \psi^{\bar b (0)}\left[ \partial_k\omega^{(0)}_{j, \bar b a}\,z^j_m z^k_{-m}
+ \partial_{\bar k}\omega^{(0)}_{\bar j, a\bar b}\,\bar z^{\bar j}_m \bar z^{\bar k}_{-m}\right],
 \eea
  with $\Omega_m = 2\pi m/\beta$. The sum $\sum'_m$ runs over all nonzero modes.
When writing this, we assumed   $m \ll M$. If $m \sim M$, one is not allowed to approximate the  finite differences in
the Euclidean action entering the finite-number-of-point approximation \p{poslepi} of the path integral
 by time derivatives.
An accurate analysis displays that the only change one should implement for large $m$  is to substitute
  \be
\lb{largem}
\Omega_m \longrightarrow  \tilde \Omega_m = -i\frac N \beta  \left( 1 - e^{-2\pi i m/N} \right)
  \ee
in Eq.\p{LE2}. However, as we will see later, this replacement affects only the overall coefficient in the
functional integral that is fixed separately, while the nontrivial dependence of the integrand on the metric
is determined by the contribution of only  first   few Fourier modes.

Thus, we keep for the moment $\Omega_m = 2\pi m/\beta$ and diagonalize the  sum in \p{LE2}  by the substitution
  \bea
\lb{zamena}
\psi^b_m  \Rightarrow  \psi^b_m  + \psi^{a (0)}[\bar z^{\bar k}_{-m}\omega^{(0)}_{\bar k, a\bar b}
- z^k_m \omega^{(0)}_{k, \bar b a}]\,, \nn
\bar\psi^{\bar b}_m  \Rightarrow  \bar\psi^{\bar b}_m  + \bar\psi^{\bar a (0)}[z^{k}_{-m}\omega^{(0)}_{k, \bar a b}
- \bar z^{\bar k}_m \omega^{(0)}_{\bar k, b \bar a}]\,.
  \eea
It brings \p{LE2} to the simple form
\be
\lb{LE22}
\int^\beta_0  L_E^{(2)} d\tau \ = \
\sum'_m A_{m a \bar b}\, \psi^a_m \bar\psi^{\bar b}_m  +
\sum'_m D_{m j \bar k}\, z_m^j\,  \bar z_{m}^{\bar k}\, ,
\ee
where
\be
A_{m a \bar b} = -i\beta \,  \Omega_m \delta_{a \bar b}\,, \quad D_{m j \bar k} =
\beta \left[ \Omega_m^2 h_{j \bar k}^{(0)}
- i \Omega_m\, R_{j\bar k, a\bar b}\,\psi^{a (0)}\bar \psi^{\bar b (0)} \right]
\ee
and $R_{j\bar k, a\bar b}$ is
 the Riemann tensor defined in (\ref{Riemann}). In the process of passing from \p{LE2} to \p{LE22} we used the
property $\Omega_{-m} = -\Omega_m$ and also the identity
\be
\lb{Rnol}
\partial_{[l}\omega_{k],\bar b a} - \omega_{[l, \bar d a}\omega_{k], \bar b d} = 0\quad ({\rm and}\;\; {\rm c.c.})
\ee
(the l.h.s. of Eq.(\ref{Rnol}) is none other than  the component
$R_{lk \bar b a}$ of the Riemann tensor that vanishes for K\"ahler manifolds).

Note that the matrix of the partial derivatives corresponding to the substitution (\ref{zamena})
is triangle and so has a unit superdeterminant. The super-Jacobian for the variable
change \p{Fourier} is also equal to unity, because the bosonic and fermion determinants cancel each other.
The functional integral over non-zero modes is then given by a product of a large (in the continuous limit, infinite) number of
 finite-dimensional determinants, which can be symbolically written as
 \bea
\lb{detprodn}
{\rm grav.\ factor} &=&
 \mu \prod_m' \prod_j  \frac {dz^j_m d\bar z^{\bar j}_m}{2\pi}  \prod_a d\psi^a_m d\bar \psi^{\bar a}_m \,
 \exp \left\{ - A_m \psi_m  \bar \psi_m - D_m  z_m \bar z_m \right\}
\nonumber \\
&=& \mu \prod'_m \det \|A_m \| \cdot {\det}^{-1} \|D_m\| \ ,
 \eea
where
\be
\lb{measure}
\mu \ =\ \left( \det \|h_{i\bar k}^{(0)} \| \right)^N \prod_{m=1}^M \Omega_m^{2n}
 \ee
 is the appropriate measure. The factor $\left( \det \|h_{i\bar k}^{(0)} \| \right)^N$ in \p{measure}
comes from the
factor $\prod_{r=0}^{N-1}\,  \det \|h_{i\bar k} (\tau_r) \|$  in \p{poslepi},
where the dependence of $z^j(\tau), \bar z^{\bar j}(\tau)$ on higher Fourier harmonics has been suppressed.
This suppression can be justified by noticing that the characteristic values of $z_m$ in
the integral $\prod_{mj} dz^j_m d\bar z^{\bar j}_m \, \exp \{ - \int_0^\beta L_E^{(2)} d\tau \} $
   are $z_m \sim 1/\Omega_m \sqrt{\beta} \sim \sqrt{\beta}$,
which is small at small $\beta$. The dimensional factor $\beta^{-2Mn}$ in  \p{measure} comes from the factor
 $\beta^{-nN} = \beta^{-2nM} \times \beta^{-n}$
in \p{poslepi} (the factor $\beta^{-n}$ having been borrowed to be displayed in the  constant mode integral \p{ordinary}
 after  performing the integration over
momenta). To derive from \p{poslepi} the correct numerical factor in the measure, notice that the coefficient $N^N$
present in
 \p{poslepi} can be represented as
  \be
 N^N \ =\ \prod_{m=1}^M (\beta \tilde \Omega_m) (\beta \tilde \Omega_{-m}) \, ,
  \ee
which follows in turn from the known identity
\footnote{Consider $P(x) = x^N -1 = \prod_{r=0}^{N-1}  (x- w^r)$ and calculate $P'(1)$.}
 \be
\lb{korni}
\prod_{r=1}^{N-1} (1 - w^r) \ =\ N\ , \ \ \ \ \ \ \ \ \ \ \ \ \ {\rm if}  \ \  w = e^{2\pi i/N} \ .
 \ee
Then, bearing in mind that only first few values $m$ are relevant (see below), we can replace $\tilde \Omega_m \to \Omega_m$,
which yields \p{measure}.
It is much easier, of course, to fix the factor in \p{measure}
 from the condition  that the r.h.s. of Eq.(\ref{detprodn}) is equal to 1 in the flat case $h_{j\bar k} = \delta_{j\bar k}\,$.

The calculation  gives
 \bea
\lb{detotvet}
{\rm grav.\ factor} &=& \left( \det \|h_{i\bar k}^{(0)} \| \right)^N
 \prod_m' \frac {\Omega_m^{2n}}{\det \|h_{i\bar k}^{(0)}\| {\det} \| \Omega_m^2 \delta_j^q - i\Omega_m R_j^q \|}\nn
&= &
 \det \|h_{i\bar k}^{(0)} \|
 \prod_{m=1}^\infty \, \frac  {\Omega_m^{2n}}{\det \|\Omega_m^2 \delta_j^q + R_j^s R_s^q \|} \ ,
 \eea
where
\be
R_j^q \ =\ h^{\bar k q} R_{j \bar k l \bar p}  \psi^{l (0)} \bar\psi^{\bar p(0)} \
 \ee
and we took into account the relation $N = 2M+1$ and sent $M \to \infty$ afterwards.
 We see that only {\it one} power of the determinant
$ \det \, h^{(0)}$ is left.

The infinite  $m$-product in (\ref{detotvet}) can be done by writing the determinant as the product of the eigenvalues and using the
identity
 \be
 \prod_{m=1}^\infty \, \frac {(2\pi m)^2}{(2\pi m)^2 + a^2} \ =\ \frac {a}{2 \sinh (a/2)}\ .
 \ee
For $a \sim 1$, only few first values of $m$ are essential in this product, and it justifies as promised
the assumption $m \ll M$ under which
Eq.\p{LE2} was derived.

We finally obtain
\be
 I^{\rm pure\ gravity} \ =\ \frac 1{(2\pi \beta)^n}
\int \prod_{j=1}^n d\bar z^{\bar j} dz^j d\psi^{j} d\bar\psi^{\bar j} \
\det \, \frac {\beta  R/2}{\sinh (\beta  R/2)} \ ,
 \ee
where we suppressed the superscripts $^{(0)}$ and passed back to the integration over the fermionic zero modes with the world
indices $\psi^j, \bar\psi^{\bar j}$ (this absorbs the remaining factor $\det h$ in \p{detotvet}).
Multiplying the integrand by $\exp\{- i\beta {\cal F}_{j \bar k} \psi^{j} \bar\psi^{\bar k } \}$
and doing the fermion integral, we arrive at \p{detlam} and hence to (\ref{indAS})
\footnote{
To establish the exact correspondence, one has to keep in mind that
the skew-symmetric matrix ${\cal R}$ defined in \p{FR} is represented in the K\"ahler case as
\be
{\cal R}^{AB} \ =\ \left( \begin{array}{cc} 0 & -R^{a\bar b} \\ R^{b \bar a} & 0 \end{array} \right),
\ee
where
\be
R^{a\bar b} = e^a_ie^{\bar b}_{\bar k}R^{i \bar k}_{\;\; j\bar t} \, dz^j\wedge d\bar z^{\bar t}\,.
\ee
}. It is clear now why, in this particular case, we had to insert the 1-loop gravitational factor
in the tree-level integral (\ref{ordinary}) for the index.  Formally, the factor \p{detotvet} tends to 1 for small
$\beta$ and, naively, the corrections involving $\beta$ and its higher powers can be neglected. We see, however, that each factor
$\beta$ in the expansion is multiplied by a bi-fermion structure $\sim \psi \bar \psi$, as is also the case for the expansion
of the integrand in \p{ordinary}. For the fermion integral not to vanish, we have to pick up the terms $\sim \beta^n (\psi\bar\psi)^n$
in the expansion of both the factor $\exp\{-i\beta {\cal F} \psi \bar \psi \}$ inside the tree-level integral and
of the 1-loop factor \p{detotvet} --- they come on equal footing.

On the other hand, the possible semiclassical corrections involving more powers of $\beta$ than those coming from $\psi \bar\psi$ are not
relevant (cf. a remark in the paragraph after \p{measure}).
This justifies neglecting two-loop and higher-loop effects in the functional integral
\p{indpathH}.

\section{Final comments and summary}
\setcounter{equation}0

In the previous Section, we have constructed the proof
of the Atiyah-Singer theorem for the standard Dirac operator based on the analysis of our SQM model for the K\"ahler manifolds
and for Abelian gauge fields. 
The same method can be and was used, however, to prove it for any even-dimensional manifold.
To this end, one should consider the system defined by the Lagrangian \p{N=1} accompanied by the external gauge field Lagrangian \p{gaugComp}.

As was discussed above, in the generic case, the ${\cal N} =2$ supersymmetry algebra
is realized not by the supercharges \p{DS}, but by the supercharges $/\!\!\!\!{\cal D}$ and $/\!\!\!\!{\cal D} \gamma^{D+1} $.
The Witten index \p{indWit} of this model still coincides with the Atiyah-Singer index of  $/\!\!\!\!{\cal D}$. One can be easily convinced
in it  by introducing the holomorphic variables $\chi^1 = \psi^{1-i2}/\sqrt{2}, \chi^2 =  \psi^{3-i4}/\sqrt{2}$, etc, and noting that
 \be
\lb{gamD+1}
 \gamma^{D+1} \equiv (2i)^{D/2} \prod_{A=1}^D \psi^A = \prod_{a=1}^{D/2} ( \bar \chi^a \chi^a -  \chi^a  \bar \chi^a) \equiv  (-1)^F \ .
  \ee
Then we have to expand the Euclidean version of the Lagrangians \p{N=1} and \p{gaugComp} into the modes and to perform basically
the same calculation as described above.
It  gives the same answer \p{indAS}. Exploring somewhat more complicated SQM systems,
this method can  be generalized to non-Abelian gauge fields too.

 For the index of the Dolbeault complex in a generic complex case
(called sometimes {\it arithmetic genus} by mathematicians), especially, in the case where the torsion
form is not closed, the life is much more difficult.
\footnote{When the torsion is closed, the calculation is still possible.
Such calculation is, in fact, the subject
of Ref. \cite{Bismut}
(even though this paper is purely mathematical and does not refer to functional integrals, etc)
and Ref.\cite{Mavr}.}
 The Lagrangian \p{1comp} involves in this case also
the large (with respect to the $\beta$ counting) 4-fermion term such that one cannot neglect the higher-loop
 contributions anymore. Starting from the complex dimension 2,
two-loop
contributions do not vanish, starting from the complex dimension 4,  one should also add 3-loop contributions, etc.
As a result, a direct evaluation of the functional integral is not possible anymore. Still,
one can obtain the integral representation for the
index (the so called {\it Hirzebruch-Riemann-Roch theorem})
by {\it deforming} the system considered in this paper in such a way that the  torsions vanish while supersymmetry
survives the deformation (see a recent paper \cite{HRR} for details).

Coming back to the calculation of Sector 6, we want to mention that
there is another way to evaluate the curvature-dependent corrections to the naive leading order semiclassical result
(\ref{indnaive}). One can proceed in the framework of the Hamiltonian formalism and notice that the index is given by the phase
space integral of the Weyl symbol of the operator $e^{-\beta H}$.  The point is that, generically,  $\left[e^{-{\beta H}} \right]_W$
differs from $e^{-\beta H_W}$ and there appear corrections involving higher powers of $\beta\,$.
The simplest correction of this type for a generic SQM system with the phase space variables $(p_j,q_j; \bar\psi_a, \psi_a)$
is expressed as \cite{supertrace}
 \be
\lb{expWeyl}
\left[e^{-{\beta H}} \right]_W \ =\ e^{-\beta H_W}\left(1 + \delta + O(\beta^4) \right)
 \ee
with
\bea
\lb{delta}
\delta(p_j,q_j; \psi_a, \bar\psi_a) &=& \frac {\beta^2}{48} \left[ \frac {\partial^2}{\partial \Psi_a \partial \bar\psi_a}
- \frac {\partial^2}{\partial \psi_a \partial \bar\Psi_a} + i \left( \frac {\partial^2}{\partial q_j \partial P_j}
- \frac {\partial^2}{\partial Q_j \partial p_j} \right) \right]^2 \nn
&& \times \,H(p_j,q_j; \bar\psi_a, \psi_a) H(P_j,Q_j; \bar\Psi_a, \Psi_a)|_{P=p, Q=q; \bar\Psi= \bar\psi, \Psi= \psi}\,.
 \eea
In most cases, this correction is suppressed at small $\beta$ and so is irrelevant. However, in our case, it gives a relevant $\beta$-independent
contribution,
 \be
\lb{betcorrindex}
\Delta_\delta I \ =\ - \frac 1{96\pi^2} \int \prod_j dz^j d\bar z^{\bar j} \, h^{{\bar l}k} h^{{\bar t}p}
\epsilon^{ms} \epsilon^{{\bar n} {\bar q}} R_{k{\bar t} m {\bar n}}  R_{p{\bar l} s {\bar q}} = - \frac \tau 8 \ ,
 \ee
where $\tau$ is the Hirzebruch signature. This coincides with the second term in (\ref{ind4}).

We see that the Lagrangian method is much more convenient than the Hamiltonian one: the one-loop correction
manifestly seen within the Lagrangian method corresponds to a complicated series in $\beta$ on the Hamiltonian side.
To find a relevant $\propto \beta^4$ term in the expansion  (\ref{expWeyl}) is already a pretty difficult task.

\vspace{1mm}

Finally, it is worth mentioning that there are also other cases when the index cannot be
expressed as the simple phase space integral (\ref{ordinary}).
First of all, this concerns the  systems with the continuous spectrum, like superconformal quantum mechanics \cite{confSQM} or
super-Yang-Mills quantum mechanics, where the integrals like (\ref{ordinary}) give meaningless fractional values \cite{SYM1}-\cite{SYM5}.
In these cases, due to the absence of the gap, such integrals cannot be ``focused'' on zero energy normalized states, but are
``contaminated'' by the states from continuum.

The systems with continuous spectrum are widely known and discussed in the literature.
  There is, however, {\it another} interesting class of systems, the  SQM systems related
to Abelian \cite{chiralSQED} and non-Abelian \cite{Blok} chiral supersymmetric $4D$ gauge
 theories. In the latter case, the spectrum seems to be discrete,
the index is well-defined, and still the integral (\ref{ordinary}) gives a fractional value. It would be rather
interesting to see whether this ``anomaly'' can be
  cured by taking into account  the 1-loop determinant  in the spirit of \p{indAS}.

There is also a problem in the index calculation for ``symplectic'' supersymmetric ${\cal N} = 4$
$\sigma$-models  with bosonic part describing the motion over a $3D$ conformally flat manifold \cite{sympl1,sympl2,sympl3}.
For example, for $S^3$, the index is equal to 2, while the integral (\ref{ordinary})
gives a meaningless irrational number. One of us has shown in \cite{supertrace} that
the corrections to (\ref{ordinary}) {\it are} present in this case and that they are of the same order as the
tree-level contribution. It would be interesting to try to sum up all such corrections by the
Lagrangian functional integral method.
\vspace{0.3cm}

As follows from the text of our paper, it involves both the review of the known facts concerning the interrelations
between the complex geometry and ${\cal N}=2,1$ supersymmetric quantum mechanics and a considerable amount of the new
results in this area. For convenience of the reader, we finish with the short summary of these new findings.\\

\noindent 1. We constructed, for the first time, quantum supercharges and Hamiltonian for the ${\cal N}=2$ SQM model
\p{start} in the case of general complex $n$-dimensional manifold and established the one-to-one correspondence
of this system with twisted and untwisted Doulbeault complexes.
\vspace{0.1cm}

\noindent 2. For the K\"ahler manifolds, we also found a correspondence
with twisted and untwisted Dirac complexes and confirmed the equivalence of the twisted Dirac and Dolbeault
complexes for this case.
\vspace{0.1cm}

\noindent 3. We presented a new detailed calculation of the index of the Dirac operator for the K\"ahler manifolds within the considered
${\cal N}=2$ SQM model and thereby gave one more ``physical'' proof of the Atiyah-Singer theorem for this operator.
\vspace{0.1cm}

\noindent 4. We presented ${\cal N}=2$ superfield formulation of  the K\"ahler ${\cal N}=4$ SQM model based on the off-shell
${\cal N}=4$ multiplet $({\bf 2, 4, 2}) = ({\bf 2,2, 0})\oplus ({\bf 0,2, 2})\,$.

\bigskip
\noindent
\section*{Acknowledgements}
We are indebted to J. Buchbinder, S. Fedoruk, M. Jardim, A. Nersessian,
V. Roubtsov, I. Smilga, S. Theisen, and, especially,
to A.Wipf
for useful discussions and correspondence.
The work of E.I. was supported in part
by RFBR grants 09-02-01209, 09-01-93107, 09-02-91349 and a grant of the Heisenberg-Landau program.
He thanks SUBATECH, Universit\'e de Nantes,
for the kind hospitality in the course of this study.


\bigskip
\bigskip


\begin{thebibliography}{99}

\bibitem{EJH} A physicist may consult [T. Eguchi, P.B. Gilkey, and A.J. Hanson,
Phys. Repts. {\bf 66} (1980) 213].

\bibitem{AS1} M.F. Atiyah and I.M. Singer, Annals Math. {\bf 87} (1968) 484.


\bibitem{AS2} M.F. Atiyah and I.M. Singer, Annals Math. {\bf 87} (1968) 546.

\bibitem{AS3} M.F. Atiyah and I.M. Singer, Annals Math. {\bf 93} (1971) 119.

\bibitem{AS4} M.F. Atiyah and I.M. Singer, Annals Math. {\bf 93} (1971) 139.

\bibitem{Zumino1}  B. Zumino, {\it Supersymmetry and the index theorem}
Preprint LBL-17972, UCB-PTH-84-17 (Lecture at the Shelter Island Conference, Shelter Island, NY, June 1983).
\bibitem{Zumino2} J. Ma$\tilde{\rm n}$es and B.Zumino, Nucl. Phys. {\bf B270} (1986) 651.

\bibitem{crp1} A.S. Schwarz, Phys. Lett. {\bf 67B} (1977) 172.
\bibitem{crp2} R. Jackiw and C. Rebbi, Phys. Lett. {\bf 67B} (1977) 189.
 \bibitem{crp3} C. Callan, R. Dashen, and D. Gross, Phys. Rev. {\bf D17} (1978) 2717.

\bibitem{LectQCD} See e.g. [A.V. Smilga, {\it Lectures on Quantum Chromodynamics},
World Scientific 2001, Chapt. 12] for detailed explanations.

\bibitem{WitMorse1} E. Witten, Nucl. Phys. {\bf B188} (1981) 513.
\bibitem{WitMorse2} E. Witten,  J. Diff. Geom. {\bf 17} (1982) 661.

\bibitem{Alv} L. Alvarez-Gaume, Commun. Math. Phys. {\bf 90} (1983) 161
\bibitem{FW} D. Friedan and P. Windey,
Nucl. Phys.  {\bf B235} (1984) 395.
\bibitem{Wind} P. Windey, Acta Phys. Polon. {\bf B15} (1984) 435.

\bibitem{Most1} A. Mostafazadeh, J. Math. Phys. {\bf 35} (1994) 1095, {\tt arXiv:hep-th/9309060}.
\bibitem{Most2} A. Mostafazadeh, J. Math. Phys. {\bf 35} (1994) 1125, {\tt arXiv:hep-th/9309061}.
\bibitem{Most3} A. Hietamaki, A.Yu. Morozov, A.J. Niemi, and K. Palo, Phys. Lett. {\bf B263} (1991) 417.

\bibitem{Hull} C.M. Hull,
{\tt arXiv:hep-th/9910028}.


\bibitem{Wipf} A. Kirchberg, J.D. Lange, and A. Wipf, Ann. Phys. {\bf 315} (2005) 467, {\tt arXiv:hep-th/0401134}.

\bibitem{arch} E.A. Ivanov and A.V. Smilga, {\tt arXiv:1012.2069v1 [hep-th]}.

\bibitem{S4} A.V. Smilga, SIGMA {\bf 7} (2011) 105, {\tt arXiv:1105.3935 [math-ph]}.

\bibitem{flux} A.V. Smilga, J. Math. Phys. {\bf 53} (2012) 042103, {\tt arXiv:1104.3986 [math-ph]}.

\bibitem{HRR} A.V. Smilga, SIGMA {\bf 8} (2012) 003, {\tt arXiv:1109.2867 [math-ph]}.

\bibitem{FIS} S.A. Fedoruk, E.A. Ivanov, and A.V. Smilga, {\tt arXiv:1204.4105 [hep-th]}.

\bibitem{Bismut} J.-M. Bismut, Math. Ann. {\bf 284} (1989) 681.

\bibitem{Mavr} N.E. Mavromatos, J. Phys. {\bf A21} (1988) 2279.


\bibitem{quantCP11}E.~D'Hoker and L.~Vinet,
  Phys.\ Lett.\  {\bf B137} (1984) 72.
\bibitem{quantCP12} V.~P.~Akulov and A.~I.~Pashnev,
  Theor.\ Math.\ Phys.\  {\bf 65} (1985) 1027
  [Teor.\ Mat.\ Fiz.\  {\bf 65} (1985) 84].
\bibitem{quantCP13} G.~A.~Mezincescu and L.~Mezincescu,
  J.\ Math.\ Phys.\  {\bf 44} (2003) 3595, {\tt arXiv:hep-th/0109002}.




\bibitem{howto} A.V. Smilga, Nucl. Phys. {\bf B292} (1987) 363.

\bibitem{Groen} H.J. Groenewold, Physica {\bf 12} (1946), pp. 405-460.
\bibitem{Moyal} I.E. Moyal, Proc. Cambr. Phil. Soc. {\bf 45} (1949) 99.

\bibitem{HKT} P.S. Howe and G. Papadopoulos, Phys. Lett. {\bf B379} (1996) 80.

\bibitem{DelIv} F.~Delduc and  E.~Ivanov, Nucl. Phys. {\bf B855} (2012) 815, {\tt arXiv:1107.1429[hep-th]}.

\bibitem{Braden} H.\,Braden, 
 Ann. Phys. NY {\bf 171} (1986) 433.


\bibitem{Konush} M.A. Konyushikhin and A.V. Smilga, Phys. Lett. {\bf B689} (2010) 95, {\tt arXiv:0910.5162 [hep-th]}.

\bibitem{IO} E.~Ivanov and O.~Lechtenfeld,
  JHEP {\bf 0309} (2003) 073, {\tt arXiv:hep-th/0307111}.

\bibitem{IKS}E.A.~Ivanov, M.A.~Konyushikhin, and A.V.~Smilga,
JHEP {\bf 1005} (2010) 033, {\tt arXiv:0912.3289 [hep-th]}.


\bibitem{Zumino} B. Zumino, Phys. Lett. {\bf 87B} (1984) 203.

\bibitem{Davis} A.C. Davis, A.J. Macfarlane, P.Popat, and J.W. van Holten, J. Phys. {\bf A17} (1984) 2945.
\bibitem{Macfar} A.J. Macfarlane and P.C. Popat, J. Phys. {\bf A17} (1984) 2955.


\bibitem{sigma1} P. di Vecchia and S. Ferrara, Nucl. Phys. {\bf B130} (1977) 93.
\bibitem{sigma2}  E. Witten, Phys. Rev. {\bf D16} (1977) 2991.
\bibitem{sigma3} D.Z. Freedman and P.K. Townsend, Nucl. Phys. {\bf B177} (1981) 282.

\bibitem{nonl1} E. Ivanov, S. Krivonos, and O. Lechtenfeld,
Class. Quant. Grav. {\bf 21} (2004) 1031, {\tt arXiv:hep-th/0310299}.
\bibitem{nonl2} S. Bellucci, S. Krivonos, A. Marrani,
and E. Orazi,
Phys. Rev. {\bf D73} (2006) 025011, {\tt arXiv:hep-th/0511249}.



\bibitem{Nicola} L.I. Nicolaescu, {\it Notes on Seiberg-Witten theory}, AMS, Providence, 2000.

\bibitem{SmHKT} A.V. Smilga, arXiv:1209.0539 [math-ph].

\bibitem{Cecotti} S. Cecotti and L. Girardello,
 Phys. Lett. {\bf B110} (1982) 39.
\bibitem{Mukhi} L. Girardello, C. Imbimbo, and S. Mukhi., Phys. Lett. {\bf B132} (1983) 69.

\bibitem{WitSYMCS} E. Witten, in
[Shifman M.A., ed.: {\it The many faces of the superworld}, World Scientific, Singapore, 2000, p.156],
{\tt arXiv:hep-th/9903005}.


\bibitem{SmSYMCS1} A.V. Smilga, JHEP {\bf 1001} (2010) 086,  {\tt arXiv:0910.0803 [hep-th]}.
 \bibitem{SmSYMCS2}  A.V. Smilga, JHEP {\bf 1205} (2012) 103, {\tt arXiv:1202.6566 [hep-th]}.


\bibitem{imt} E. Ivanov, L. Mezincescu, and P.K. Townsend,
in [Salamanca 2003, {\it Symmetries in gravity and field theory}, p.385], {\tt arXiv:hep-th/0311159}.



\bibitem{supertrace} A.V. Smilga, Commun. Math. Phys. {\bf 230} (2002) 245, {\tt arXiv:hep-th/0110105}.

\bibitem{confSQM} S. Fubini and E. Rabinovici, Nucl. Phys. {\bf B245} (1984) 17.

\bibitem{SYM1} A.V. Smilga, Nucl. Phys. {\bf B266} (1986) 45.
\bibitem{SYM2} P. Yi, Nucl. Phys. {\bf B505} (1997) 307, {\tt arXiv:hep-th/9704098}.
\bibitem{SYM3} S. Sethi and M. Stern, Commun. Math. Phys. {\bf 194} (1998) 675, {\tt arXiv:hep-th/9705046}.
\bibitem{SYM4} G. Moore, N. Nekrasov, and S. Shatashvili,
Commun. Math. Phys. {\bf 209} (2000) 77, {\tt arXiv:hep-th/9803265}.
\bibitem{SYM5} V.G. Kac and A.V. Smilga,
Nucl. Phys. {\bf B571} (2000) 515, {\tt arXiv:hep-th/9908096}.


\bibitem{chiralSQED} A.V. Smilga,  JETP {\bf 64} (1986) 8.

\bibitem{Blok} B. Yu. Blok and A.V. Smilga, Nucl. Phys. {\bf B287} (1987) 589.



\bibitem{sympl1} A.V. Smilga, Nucl. Phys. {\bf B291} (1987) 241.
\bibitem{sympl2} E.A. Ivanov and A.V. Smilga,
Phys. Lett. {\bf B257} (1991) 79.
\bibitem{sympl3}  V.P. Berezovoj and A.I. Pashnev, Class. Quant. Gravity {\bf 8} (1991) 2141.


\end{thebibliography}
\end{document}